\documentclass[11pt,letterpaper]{article}
\pdfoutput=1

\usepackage{graphicx,array}
\usepackage[dvipsnames,table,xcdraw]{xcolor}
\definecolor{darkblue}{rgb}{0,0.1,0.5}
\definecolor{darkgreen}{rgb}{0,0.5,0.2}
\definecolor{darkred}{RGB}{153,26,0}
\usepackage{ulem}
\usepackage{slashed}
\definecolor{seablue}{rgb}{0,0.2,0.6}
\usepackage{latexsym}
\usepackage{amssymb, amsmath}
\usepackage{slashed}
\usepackage{bm}        
\usepackage[numbers,sort&compress]{natbib}
\usepackage{bm,relsize}
\usepackage{slashed}
\usepackage{tcolorbox}
\definecolor{viola}{RGB}{134,41,198}
\usepackage{mathrsfs}
\usepackage{hyperref} 
\hypersetup{
    colorlinks=true,       
    linkcolor=darkblue,          
    citecolor=BrickRed,        
    filecolor=magenta,      
    urlcolor=MidnightBlue           
}
\usepackage[all]{hypcap} 
\usepackage{subcaption}

\usepackage{tikzfeynman}

\usepackage[font=small,labelfont=bf,labelsep=period]{caption}

\newcommand{\lin}{\langle \mathrm{in}|}
\newcommand{\rin}{|\mathrm{in}\rangle}

\newcommand{\rout}{|\mathrm{out}\rangle}

\newcommand{\te}{{\bf e}}

\usepackage{stackengine}
\stackMath

\newcommand{\lp}{\left(}
\newcommand{\rp}{\right)}

\newcommand{\be}{\begin{equation}}
\newcommand{\ee}{\end{equation}}

\definecolor{lightergray}{rgb}{0.9,0.9,0.9}

\begin{document}

\begin{flushright}

\end{flushright}
\vspace{-1cm}
\begin{center}
{\LARGE \bf 
Particle production from inhomogeneities:\\
general metric  perturbations 
}\\
\bigskip\vspace{1cm}
{
\large Raghuveer Garani$^{a}$, Michele Redi$^b$, Andrea Tesi$^b$}
\\[7mm]
 {\it \small
$^a$ Department of Physics, Indian Institute of Technology Madras, Chennai 600036, India\\
$^b$INFN Sezione di Firenze, Via G. Sansone 1, I-50019 Sesto Fiorentino, Italy\\
Department of Physics and Astronomy, University of Florence, Italy
}
\end{center}

\vspace{.2cm}

\centerline{\bf Abstract} 
\begin{quote}
We present universal formulas for particle production from gravitational inhomogeneities. 
In the massless limit the result is strikingly simple and completely determined by the two-point function 
of the energy-momentum tensor that is fixed up to a constant -- the central charge -- for conformally coupled scalars, massless fermions and gauge fields. 
This result can be applied to any conformally coupled theory, weakly or strongly interacting, unifying 
previous derivations for fields of different spin and for scalar and tensor perturbations. 
We derive the results using the Schwinger method of 1PI effective action and through Bogoliubov transformations
that allows to compute exclusive information on the distribution of particles.  
We then apply these results to stochastic backgrounds of scalar and tensor perturbations that can be generated by 
various phenomena such us inflationary perturbations and first order phase transitions. 
Differently from particle production usually considered in cosmology this mechanism allows for the production of massless fields. 
In particular the abundance induced by inhomogeneities can easily reproduce the dark matter abundance if scalar perturbations produced 
from inflation are enhanced at short scales. 
\end{quote}

\vfill
\noindent\line(1,0){188}
{\scriptsize{ \\ E-mail:\texttt{  \href{mailto:garani@iitm.ac.in}{garani@iitm.ac.in}, \href{mailto:michele.redi@fi.infn.it}{michele.redi@fi.infn.it}, \href{andrea.tesi@fi.infn.it}{andrea.tesi@fi.infn.it}}}}
\newpage

\tableofcontents

\section{Introduction}

A very minimal and unavoidable production mechanism for dark matter (DM) is gravitational particle production (GPP) \cite{Ford:1986sy}. This happens since 
in the homogeneous and isotropic expanding universe the vacuum state is not well defined and particles can be produced thanks to energy non-conservation and quantum mechanics. This process can be understood as a Bogoliubov transformation between the initial and final vacuum and has been extensively studied in the literature, see \cite{Kolb:2023ydq} for a review.  The key element of the mechanism is  breaking of Weyl invariance, i.e. invariance under local rescalings the metric, $g_{\mu\nu}(x) \to \Omega^2(x) g_{\mu\nu}(x)$ that allows the expansion of the universe to come to light. The breaking can happen either because of kinetic terms that violate Weyl symmetry or 
due to mass terms. In the first case -- that is realized by minimally coupled scalars such as the inflaton and gravity waves -- the fields can be 
produced in the massless limit while in the latter the mass must be relevant so that the production takes place when the Hubble parameter 
becomes comparable to the mass. This leads to very different scenarios where DM can be either very light being copiously produced
during inflation, or very heavy when it is produced during reheating or radiation domination. 

Recently the possibility to produce DM also from metric inhomogeneities was considered in \cite{kopp,kopp2,Garani:2024isu}.
This mechanism is qualitatively different from GPP described above because it works in the limit of exact Weyl invariance 
of the action, allowing to produce conformally coupled fields such as massless fermions and gauge fields, which are notoriously not produced by the background expansion of the Universe. 
The production of fermions from stochastic scalar perturbations and  gravity waves backgrounds 
was computed with different methods, in-in formalism and Bogoliubov transformations. The abundance of particles 
is determined by the  power spectrum of the perturbations $\Delta(q\,,q_0)$, a function of comoving momenta $|\vec q|$ and time, $q_0$ being the time Fourier transformed variable. A large background of scalar perturbations can be 
for example produced during inflation. For a signal peaked at $q_*$ one finds the numerical abundance goes like $n\propto q_*^3 \Delta(\vec q_*,q_0)$ so that
the mechanism is particularly efficient if the power spectrum is large for short wavelengths. Inhomogeneities could be produced during first order phase transitions 
or other violent events in the early universe. 

In this paper we revisit and significantly extend  Ref. \cite{Garani:2024isu} to fields of any spin and even interacting Conformal Field Theories (CFT)
produced by arbitrary metric perturbations. For a stochastic background we will prove that in complete generality that the abundance of particles produced by inhomogeneities is given by
\begin{equation}
\frac{d(n a^3)}{dq_0}= \frac {c_J} {320\pi^2}   \int d(\log q) \Delta(\vec q\,,q_0) f_S(q_0\,,\vec q) \theta(q_0^2-\vec q^{\, 2})\,,
\label{eq:general}
\end{equation}
where $f_{\rm S}(q_0\,,\vec q\,)=q^4/3$ for scalar perturbations and $f_{\rm GW}(q_0\,,q)=(q_0^2-q^2)^2/2$ 
for gravity wave perturbations and $q_0$ is the energy of the particle pair produced. Here $c_J$ is the central charge of the CFT that controls the 2-point function of the energy momentum tensor. 
The emergence of the central charge follows from the fact that gravitational perturbations induce a contribution to the lagrangian $\delta L = h_{\mu\nu}T^{\mu\nu}/2$. The result has strong similarities with gravitational freeze-in where the number of particles is also determined by the central charge through the optical theorem  \cite{Redi:2020ffc}.

The formula above describes the production from inhomogeneities of conformally coupled scalars, massless Weyl fermions and massless spin-1 particles where $c_0=4/3$, $c_{1/2}=4$, $c_1=16$ and also of arbitrary CFTs. Strikingly, the formula above captures many results derived with different methods \cite{Zeldovich:1977vgo,Calzetta:1986ey,Campos:1991ff,Cespedes:1989kh,Maroto1,Maroto2,Hu:2020luk,Bassett:2001jg}.
We will prove eq. (\ref{eq:general}) in two ways, through the Schwinger method of vacuum persistence using the 1PI effective action and through Bogoliubov transformations.
The latter method is more general providing exclusive information on the distribution of particles produced  but the elegance and simplicity of Schwinger formalism is unbeatable. 
Along the way we will also determine production of particles by localized perturbations such as astrophysical objects.

We foresee a  pletora of applications of our results in cosmology and astrophysics.
As an example we consider the scenario where DM is produced by curvature perturbations that originate during inflation, extending the discussion of Ref. \cite{Garani:2024isu}.
We find that realistic DM scenarios emerge especially if the power spectrum is enhanced for the modes that exit the horizon towards the end
of inflation. Similarly first order phase transitions could produce significant amounts of particles both from scalar and gravity wave inhomogeneities.
Production by astrophysical sources and non conformally coupled fields such as gravity waves and minimally coupled scalars will also be studied elsewhere.

The paper is organized as follows. In section \ref{sec:inout} we derive particle production of conformally coupled fields using the Schwinger 
method of vacuum persistence. This only requires the knowledge of the two point function of the energy momentum tensor and leads immediately to (\ref{eq:general})
for CFTs.  In section \ref{sec:bogo} we derive differential formulas for particle production from arbitrary metric perturbations of scalars, fermions and gauge fields. 
We then show that this result can be mapped to the effective action computation integrating over phase space. In section \ref{sec:stochastic} we specialize to stochastic backgrounds
presenting universal formulas for production from scalar and gravity wave backgrounds. We apply this formalism to production from scalar perturbations from inflation in section \ref{sec:inflation}.
We conclude in \ref{sec:conclusions} with an outlook on future work and collect useful formulas in the appendices.

\section{Particle production from the effective action}
\label{sec:inout}

One slick way to compute particle production is to use the 1PI effective action as a function of external fields.
We will be mostly interested in metric perturbation in an FRLW spatially flat universe.
The most general perturbation can be parametrised as,
\be\label{eq:metric}
ds^2= a^2(\tau)[ \eta_{\mu\nu}+ h_{\mu\nu}(\tau\,, \vec x)] dx^\mu dx^\nu\,,
\ee
where the scale factor is $a(\tau)$ is determined by the Hubble equation $H^2\equiv a'^2/a^4=\rho/(3M_{\rm Pl}^2$).

We will focus on theories classically invariant under a Weyl transformation of the metric $g_{\mu\nu}(x)\to \Omega^2(x,\eta) g_{\mu\nu}(x)$.\footnote{Weyl invariance forbids mass terms so our formulas will be mostly in the massless limit. For phenomenological appllications however the results capture production of particles in the relativistic limit where the mass is negligible.}
Notably this includes massless fermions, gauge fields, and conformally coupled scalars.
For Weyl invariant theories we can remove the scale factor at the classical level redefining the fields of the theory.\footnote{
At the quantum level Weyl invariance is broken by anomalies but this does not affect the computation to leading order in the perturbations considered here.}
We can thus work directly in flat space with general metric perturbations and use flat space methods throughout.
This allows to study the problem of particle production similarly to Schwinger pair production from vacuum persistence probability 
with the notable simplification that the computations is  perturbative and fixed by the symmetries. 

As first discussed in Schwinger seminal work \cite{Schwinger:1951nm},  the vacuum persistence probability can be related to the 1PI effective action. In an ``in-out'' calculation, where $|{\rm in}\rangle$ and $|{\rm out}\rangle$ are the vacuum state in the far past and far future respectively, we have
\begin{equation}\label{persistence}
|\langle {\rm out}|{\rm in}\rangle|^2 = e^{-2 {\rm Im} \Gamma_{\rm 1PI}}\,.
\end{equation}
Here $\Gamma_{\rm 1PI}$ the 1PI effective action computed from vacuum diagrams. For a stable Lorentz invariant ground state $\Gamma_{\rm 1PI}$ is real so that the sum of vacuum diagrams is just an unobservable phase. In the presence of perturbations however an imaginary part develops that describes the decay probability. Indeed we can interpret the imaginary part of the effective action as the total number of vacuum decays, see for example \cite{Watkins:1991zt},
\begin{equation}
N_{\rm dec}=2{\rm Im} \Gamma_{\rm 1PI}\,.
\end{equation}

Technically, and in very  general terms, what we are going to discuss can be applied to any Lorentz invariant theory perturbed by
\begin{equation}\label{eq:int-O}
\Delta S = \int d^4 x \phi(x) {\cal O}(x) \,.
\end{equation}
Here $\phi(x)$ is an external (background) field that acts as a source for ${\cal O}(x)$, a local operator of the theory (which need not to be necessarily scalar). To second order in $\phi(x)$ the effective action is simply given by
\begin{equation}\label{eq:eff-general}
\Gamma_{\rm 1PI}=\frac 1 2 \int d^4 x d^4 y \phi(x)\langle  {\cal O}(x) {\cal O}(y) \rangle \phi(y)=\frac 1 2 \int \frac {d^4q}{(2\pi)^4} \phi(-q)\langle  {\cal O}(q)  {\cal O}(-q) \rangle \phi(q)\,,
\end{equation}
where the expectation value is of time ordered fields. With abuse of notation here and in the rest of the paper we denote with the same symbol field and its Fourier transform.\footnote{Our Fourier transform conventions are 
\be
\int \frac{d^4p}{(2\pi)^4} e^{-i p\cdot x},\quad\quad \int d^4x e^{i p\cdot x}
\ee
and we define
\be
\langle T {\cal O}(x) {\cal O}(y) \rangle = \int \frac{d^4p}{(2\pi)^4} e^{-i p \cdot (x-y)}\langle T {\cal O}(p) {\cal O}(-p) \rangle
\ee}
Here $q$ is a four-vector. The above expression for the 1PI effective action is valid in general, both for massive and massless theories, but as we will see it becomes extremely simple in the massless limit where the two point function 
is fixed by conformal invariance.\footnote{We can apply this formula to QED where the relevant coupling is $A_\mu(x) J(x)$. The two point function of the
current is given by $\langle \tilde J_\mu(q) \tilde J_\nu(-q)\rangle = (q^2 \eta_{\mu\nu} -q_\mu q_\nu) \Pi(q^2)$ so that the effective action obtained integrating out the fermions 
reads $\Pi(q^2) F_{\mu\nu}(q)F^{\mu\nu}(-q)$. In the massless limit $\Pi(q^2)=c \log(-p^2)$ so that the imaginary part is simply $\pi c \theta(p^2) F^2$.  Note that $c$ is the coefficient of the 
1-loop $\beta-$function. For a constant electric field  this correctly reproduces the result of Schwinger pair production. 
Remarkably this result can be obtained through an educated perturbative computation despite the fact that the Schwinger mechanism is strictly non-perturbative. }

\bigskip

Let us now consider the main focus of this paper, i.e. gravitational inhomogeneities. What will be the 1PI effective action in the background of the perturbations? In this case the external field is actually given by the metric perturbations themselves, and their coupling to the matter content is fixed by gravity. Indeed they couple to the energy momentum tensor of the theory,
\begin{equation}
\Delta S_{\rm}=\frac 1 2 \int d^4x  h_{\mu\nu}(x) T^{\mu\nu}(x)\,.
\end{equation}
The analogy with eq.~\eqref{eq:int-O} is strong. Therefore we can immediately reproduce the above argument. To quadratic order the effective actions is given by,
\begin{equation}
\Gamma_{\rm 1PI} =\frac 1 8 \int \frac{d^4q}{(2\pi)^4} \frac{d^4q'}{(2\pi)^4}h^{\mu\nu}(-q) \langle  T_{\mu\nu}(q)   T_{\rho\sigma}(q')\rangle h^{\rho\sigma}(-q')\,.
\label{eq:G1PI}
\end{equation}
Again, as in \eqref{eq:eff-general}, here the expectation value is of time ordered fields since we consider in-out states in flat space. Knowing the time-ordered 2-pt function of stress-energy tensor we can compute the effective action on any background. In general such correlator might be difficult to compute. However, a major simplification arises for CFTs where the 2-point function is fixed by conformal invariance. 
For the energy momentum tensor in 4D CFTs one finds (see for example \cite{Gubser:1997se})
\begin{equation}
\langle  T_{\mu\nu}(q)  T_{\rho \sigma }(q')\rangle =(2 \pi)^4 \delta^4(q+q') \frac{c_J}{7680 \pi^2}\Pi_{\mu\nu\rho\sigma}(q)   \log(-q^2)\,,
\label{eq:2pointTT}
\end{equation}
where
\begin{equation}
\Pi_{\mu\nu\rho\sigma}(q) \equiv \left(2\pi_{\mu\nu} \pi_{\rho\sigma}-3\pi_{\mu\rho} \pi_{\nu\sigma}-3\pi_{\mu\sigma} \pi_{\nu\rho}\right)\,,\quad\quad \pi_{\mu\nu}\equiv\eta_{\mu\nu}q^2-q_\mu q_\nu\,.
\end{equation}
The structure of the two point function is completely determined by conformal invariance up to the constant $c_J$ that is known as the central charge of the CFT.
In our normalization,  $c_0=4/3$, $c_{1/2}=4$ and $c_1=16$ respectively for conformally coupled scalars, Weyl fermions and massless gauge fields. However, our derivation also applies to the class of interacting CFTs, since all is needed is just $c_J$.

Given eq.~\eqref{eq:2pointTT} is possible to evaluate eq.~\eqref{eq:G1PI}, and since we are interested in the imaginary part of the action, we notice that the logarithm in eq.~\eqref{eq:2pointTT} develops an imaginary part for positive $q^2=q_\mu q^\mu$. Using ${\rm Im} \log x= \pi \theta(-x)$ so that the number of decays is given by
\begin{tcolorbox}[colback=lightergray, colframe=white, boxrule=0pt, left=0mm, right=0mm]
\begin{equation}
N_{\rm dec}\equiv 2{\rm Im}\Gamma_{\rm 1PI} =\frac {c_J}{30720 \pi}  \int \frac{d^4q}{(2\pi)^4} \theta(q^2)  h_{\mu\nu}(q) \Pi^{\mu\nu\rho\sigma}(q) h_{\rho\sigma}(-q)\,.
\label{eq:Nparticles}
\end{equation}
\end{tcolorbox}
At weak coupling each decay produces two particles so that the number of particles produced is just twice this.
Up to the $\theta(q^2)$ the integrand is a 4 derivative operator quadratic in the metric.  This formula is very general. The result suggests that $q_0$ should be interpreted as  the energy of the particle produced but this statement cannot be proven from the effective action alone.  As we will see this expectation is correct, even though $q_0$ is more precisely the energy of the pair of particles produced. In light of $\theta(q^2)$ a perturbation of momentum $\vec q$ can only produce pair of particles with $q_0>|\vec q|$. Nevertheless as we will see the energy spectrum of single particles extends to zero momentum. 

\medskip

Our calculation can stop here at eq.~\eqref{eq:Nparticles}, since we have already anything need to compute the abundance of conformally coupled matter in cosmology. It is however inspiring to rewrite the expression in eq.~\eqref{eq:Nparticles} by doing a step more and explicitly use the expression of the tensor $\Pi_{\mu\nu\rho\sigma}$. One can check that the expression above can be re-written as
\begin{equation}
N_{\rm dec}= \frac{c_J}{2560 \pi}  \int \frac{d^4q}{(2\pi)^4} \theta(q^2) W^{\mu\nu\rho\sigma}(q) W_{\mu\nu\rho\sigma}(-q)\,,
\label{eq:Nparticles2}
\end{equation} 
where $W_{\mu\nu\rho\sigma}(q)$ is the Fourier transform of the linearised Weyl tensor of the metric perturbations.
Therefore particle production is fully determined by the geometry and central charge.
The appearance of the Weyl tensor is due to the fact that, at 4 derivates order, $W^2_{\mu\nu\rho\sigma}$ is the only Weyl invariant tensor
that can be generated by conformally coupled fields. More in general for particles that are not conformally coupled, for example minimally coupled scalars or gravitons,
the result involves $R^2$. For example for minimally coupled massless scalars the integrand includes a term $ 60\lp\xi -\frac16\rp^2 R(q) R(-q)$~(see, eq. (5.118) in~\cite{Birrell:1982ix}).

\section{Particle Production from Bogoliubov transformations}
\label{sec:bogo}

We now discuss how to derive particle production using the formalism of Bogoliubov transformations.
This will provide more general differential formulas shedding some light on the result found using the effective action.

\subsection{Particle production for conformally coupled matter} 
To set the stage  let us first consider a conformally coupled scalar field $\phi$ of mass $M$ in the expanding universe
\begin{equation}
S=\frac 1 2 \int d^4 x \sqrt{-g}\left[ g^{\mu\nu} \partial_\mu \phi \partial_\nu \phi -\left(M^2-\frac R 6\right) \phi^2\right ]\,.
\label{eq:S0}\ee
We assume that $\phi$ evolves in the background of eq.~\eqref{eq:metric}. We are going to discuss two relevant limits. First, the case where $M\neq 0$ and $h_{\mu\nu}=0$ which corresponds to the usual GPP. Second, the case where $M=0$, such that the matter action is Weyl invariant, and $h_{\mu\nu}\neq 0$, which is new and the main focus of our paper.

\bigskip

Let us note that the Bogoliubov calculation is again related to discussing the properties of the $|\rm in\rangle$ and $|\rm out\rangle$ vacuum states in the far past and far future as in the Schwinger's approach. Here one exploits the possibility to expand the field upon a set of well defined operators in the far past ($a$) and future ($\bar{a}$) such that they annihilate the respective vacuum states
\be\label{eq:bogo-definition-vacuum}
a_{\vec k} \rin = \bar{a}_{\vec q}\rout = 0\,.
\ee
Finding the (linear) transformation between $(\bar{a}_{\vec q},\bar{a}^\dag_{\vec q'}) \leftrightarrow (a_{\vec k},a^\dag_{\vec k'})$ allows us to compute observables. Such a linear mapping is called a Bogoliubov transformation. Knowing the mapping, it is easy to show the equivalence between the two methods (see Appendix \ref{sec:equivalence}). The starting point, i.e. the $\rin$ state, fixes the initial conditions for the field in the usual way: upon some field redefinitions the field is just an infinite sum of positive frequency solutions $e^{-i k\tau}$ associated to the operator $a_{\vec k}$.

\paragraph{1. Gravitational Particle Production: $M\neq0$, $h_{\mu\nu}=0$.}~\\
In this case, which is just FRLW geometry, upon rescaling the field $\phi=\varphi/a$, the equation of motion  reads,
\begin{equation}
\square \varphi + M^2 a(\tau)^2 \varphi=0\,,
\label{eq:bogoex}
\end{equation}
where $\Box$ is the flat space D'Alembertian so that the scale factor only enters through the mass term.  

Solving the equation of motion one finds that a positive frequency wave at early times where the mass term is negligible, i.e. $\varphi \propto e^{-i k \tau}$, 
evolves into a combination of positive and negative frequencies. Since the positive/negative frequencies in the quantum theory are associated to creation and annihilation operators
this defines a Bogoliubov transformation between two isomorphic Hilbert spaces at early and late times. Because the FLRW background is  invariant under spatial translations the Bogoliubov transformation does not mix modes of different momentum. The Bogoliubov transformation in this case reads
\begin{equation}
\bar{a}_{\vec k}= \alpha_{\vec k}\,  a_{\vec k} +\beta_{\vec k}^*\, a^\dag_{-\vec k}\,,
\end{equation}
where $|\alpha_{\vec k}|^2-|\beta_{\vec k}|^2=1$ guarantees that canonical commutation relations are satisfied.
The coefficients of $\alpha_{\vec k}$ and $\beta_{\vec k}$ can be directly extracted from the solution of the wave equation (\ref{eq:bogoex}).
Assuming positive frequency at $\tau=-\infty$ the asymptotic solution at late times is given by
\begin{equation}
\varphi_{\vec k}(\tau)\sim \frac  {\alpha_{\vec k}}{\sqrt{2\omega_{\vec k}(\tau)}} e^{-i \int^\tau d\tau' \omega_{\vec k}(\tau')} +\frac  {\beta_{\vec k}}{\sqrt{2\omega_{\vec k}(\tau)}} e^{i \int^\tau d\tau' \omega(\tau')} 
\end{equation}
where $\omega_{\vec k}^2={\vec k}^2+M^2 a^2$.
The  Bogoliubov coefficient $\beta_{\vec k}$ is then the coefficient of the negative frequency wave. Note that the mass term in eq. (\ref{eq:bogoex}) becomes large at late times so that the solution must be found to all orders in $\tau$. Moreover no particle production occurs when $M=0$ because the wave equation reduces to the one of flat space. This phenomenon follows from Weyl invariance of the action that guarantees that no particles can be  produced in  conformally flat metric. The same applies to massless fermions, gauge fields and more in general CFTs where Weyl invariance, i.e. $T_\mu^\mu\equiv 0$ follows under broad 
assumptions \cite{Luty:2012ww}.

\bigskip
\paragraph{2. Inhomogeneities: $M=0$, $h_{\mu\nu}\neq 0$.}~\\
If  the background breaks translation invariance modes of different momentum can mix, see \cite{Ford:2021syk} for a nice review.
As consequence the Bogoliubov coefficients become matrices,
\begin{equation}
\bar{a}_{\vec k}= \sum_{\vec p}\alpha_{\vec k \vec p}  a_{\vec p} +\beta_{\vec k \vec p}^* a^+_{-\vec p}\,,\quad\quad\quad\sum_{\vec p} \alpha_{\vec k \vec p}\alpha_{\vec k' p}^*-\beta_{\vec k \vec p}\beta_{\vec k' \vec p}^*=\delta_{\vec k \vec k'}\,.
\label{eq:bogomatrix}
\end{equation}
On the other hand the situation is simpler than $M\ne 0$ because the perturbation term is never large and thus one can work in perturbation theory. 

Concretely to see how this works let us consider a scalar in flat space with a space-time dependent mass term  $\frac12M^2(x) \varphi^2$ that vanishes at early and late times. 
While this is much simpler than considering the metric perturbations $h_{\mu\nu}$ for conformally coupled fields, it shows the main features of the mechanism.
The equation of motion is simply
\begin{equation}\label{toy}
\square \varphi = M^2(x) \varphi\,.
\end{equation}
Before solving this equation, let us notice that in the asymptotic past and future the solutions are well known and we dub them as $\varphi^{\rm in}$ and $\varphi^{\rm out}$. In particular the Fourier components of the quantum fields are
\be
\varphi^{\rm in}_{\vec k}(\tau\to -\infty)=  \frac{e^{-i k\tau}}{\sqrt{2k}}a_{\vec{k}} + \frac{e^{i k\tau}}{\sqrt{2k}}a^\dagger_{-\vec k}\,,\quad
\varphi^{\rm out}_{\vec k}( \tau\to \infty)= \frac{e^{-i k\tau}}{\sqrt{2k}}\bar{a}_{\vec k} + \frac{e^{i k\tau}}{\sqrt{2k}}\bar{a}^\dagger_{-\vec k}
\ee
where $a_{\vec k}$ and $\bar{a}_{\vec k} $ annihilate $\rin$ and $\rout$ respectively as in \eqref{eq:bogo-definition-vacuum}. 

It is instructive to see how we can extract the Bogoliubov coefficient in this case. Given that the perturbation is a small effect,
the relation between the $a_{\vec k}$ and $\bar{a}_{\vec k}$ can be found perturbatively solving the equation for $\varphi$ with initial condition $\varphi^{\rm in}$.  Working at first order in the perturbation we get
\begin{eqnarray}
\label{eq:pert}
\varphi^{\rm out}_{\vec k}( \tau)&=&\varphi^{\rm in}_{\vec k}( \tau) + \int d\tau' G^{\rm ret}_{\vec k}( \tau-\tau')J_{\vec k}(\tau')\,,\\
 J_{\vec k}(\tau')&\equiv&\int \frac{d^3 q}{(2\pi)^3} M^2(\tau', \vec q) \varphi^{\rm in}_{\vec k- \vec q}(\tau')\,.
\label{eq:green}
\end{eqnarray}
We have introduced a source  $J_{\vec k}$, which depends upon the spatial Fourier transform of the perturbation $M^2$ and on the initial quantum field. 
The structure is such that the initial field $\varphi^{\rm in}$ is evaluated at the shifted momentum $\vec k-\vec q$. For what follows we then define,
\begin{equation}
\vec \omega\equiv \vec k -\vec q\,,~~~~~~~\omega \equiv |\vec k -\vec q|= \sqrt{k^2+q^2- 2 q k \cos \theta}\,.
\end{equation}
The retarded Green's function for scalars is given by,
\begin{equation}
G^{\rm ret}_{\vec k}( \tau)= \theta(\tau) \bigg[ \frac{i}{2k}e^{-i k \tau}-\frac{i}{2k}e^{i k \tau}\bigg]\,,
\end{equation}
so that (\ref{eq:pert}) automatically produces a combination of positive and negative frequency waves. Equating the coefficients of the exponential terms, we can immediately 
identify the explicit relation between the two set of in and out operators

\begin{equation}\label{eq:a-bar}
\bar a_{ \vec k}= \bar{a}_{\vec k}-i\int \frac{d^3q}{(2\pi)^3}\int d\tau' \left[\frac{e^{i (k-\omega)\tau'}}{2\sqrt{k \omega}}M^2(\tau',\vec q) a_{\vec \omega} +\frac{e^{i (k+\omega)\tau'}}{2\sqrt{k \omega}}M^2(
\tau',\vec q)  a^\dag_{ -\vec \omega}\right]\,.
\end{equation}
From this example we see that the out operators can be expressed as a linear combination of the initial ones. 
We are then led to define, more generally, the following linear transformation that maps the in operators onto the out operators,
\be
\left(\begin{array}{c} \bar{a}_{\vec k} \\ \bar{a}^\dag_{-\vec k}\end{array}\right)= \int \frac{d^3q}{(2\pi)^3}\left(\begin{array}{cc} \alpha_{\vec k, \vec \omega} & \beta^*_{\vec k, \vec \omega} \\ \beta_{-\vec k, -\vec \omega} & \alpha^*_{-\vec k, -\vec \omega} \end{array}\right)\left(\begin{array}{c} a_{ \vec \omega} \\ a^\dag_{ -\vec \omega}\end{array}\right)
\label{eq:bogorotation}
\ee
Here $\alpha_{\vec k \vec \omega}$ and $\beta_{\vec k \vec \omega}$ are generalized Bogoliubov coefficients.
Roughly speaking, this is just the continuous version of (\ref{eq:bogomatrix}).

Let us inspect directly the expression of $\beta_{\vec k\vec \omega}$ in this example, since it is all we need to compute observables. Explicitly in this case we have

\be
\begin{split}
\beta^*_{\vec k, \vec \omega}&=-\frac i {2\sqrt{k \omega}}\int d\tau' e^{i (k+\omega)\tau'}M^2(\tau',\vec q)\equiv -\frac i {2\sqrt{k \omega}}M^2(k+\omega,\vec q)\,, \\
\alpha_{\vec k, \vec \omega}&=(2\pi)^3 \delta^3(\vec k-\vec q) -\frac i {2\sqrt{k \omega}}M^2(k-\omega,\vec q)\,.
\end{split}
\ee

As we will see a similar expression holds in more complicated situations involving derivatives of the perturbation using integration by parts
(boundary terms vanish at plus and minus infinity by constructions). Formulas becomes extremely simple in Fourier transform as seen above.

\bigskip

From the Bogoliubov coefficients we can compute the quantities of interest. 
In particular the (differential) particle number of and energy is given in the massless limit
\begin{equation}
\frac {dN}{dk d(\cos \theta_k) d\phi_k}=  \frac {k^2}{(2\pi)^3} \int \frac{d^3 q}{(2\pi)^3} \left|\beta_{\vec k\vec \omega}\right|^2\,, \quad\quad \frac {dE}{dk d(\cos \theta_k) d\phi_k}=  \frac {k^2}{(2\pi)^3} \int \frac{d^3 q}{(2\pi)^3} k 
\left|\beta_{\vec k\vec \omega}\right|^2\,.
\label{eq:Ndiff0}
\end{equation}
~\\
Here the angles $(\theta_k\,,\phi_k)$ label the direction of the produced particles and $k$ is their energy. The derivation of these formulas is carried out in the appendix \ref{app:bogo}. At this level we already see how the Bogoliubov derivation allows to compute more exclusive observables than the Schwinger's effective action approach.
Completely analogous formulas apply to particles with spin. Eq.s~\eqref{eq:Ndiff0} are general - despite being derived for our example - and we will use them throughout the following section.

\section{Production from metric inhomogeneities}

We now wish to generalise the discussion above to  metric perturbations around the expanding FLRW background parametrized by eq. (\ref{eq:metric}).
As we mentioned for conformally coupled fields, Weyl invariance of the action implies that the scale factor can be removed from the equations of motions. As a consequence we can directly work around flat space with $a(\tau)=1$. In particular, in this section, we cover the following cases
\begin{itemize}
\item \textbf{Massless conformally coupled scalar}.~\\ 
This is the case of the previous section, and we have seen explicitly that the Weyl rescaling acts as $\varphi \to a^{-1} \varphi$
\item \textbf{Massless fermion}.~\\ We consider a massless Weyl fermion, described by the following action
\begin{equation}\label{eq:Sfermion}
S_{\frac 1 2} =\int d^4x \sqrt{-g} i \bar{\psi} {\bar \sigma}^a e_a^\mu (\partial_\mu+\omega_\mu) \psi 
\end{equation} 
where $\bar{\sigma}^a\equiv (1, -\vec{\sigma})$ and $\omega_\mu$ is the spin connection, see appendix for notation and further details. 
For such a field the Weyl rescaling is  $\psi \to a^{-3/2} \psi$.
\item \textbf{Gauge field}. Massless spin-1 vectors (gauge fields) have Weyl weight equal to zero, and their action is given by
\begin{equation}
\label{eq:Svector}
S_1=-\frac 1 4 \int d^4 x \sqrt{-g}  g^{\mu\alpha} g^{\nu\beta} F_{\mu\nu} F_{\alpha\beta}
\end{equation}
Therefore there is no Weyl rescaling needed,
$A_\mu \to A_\mu$.
\end{itemize}
In all cases after the Weyl rescaling, the field satisfy $\Box \phi = J(x)$, where $\Box$ is the D'Alembertian in flat space and the source is a bilinear in the metric perturbation and the field itself. 
This equation can be easily solved to first order in perturbations using retarded Green's functions.

\subsection{Master formulas}

We now present explicit formulae for scalar, fermions and gauge fields. 

The procedure is the following: $i)$ We evaluate the source term on the solution of the unperturbed equations, that is the ``in" solution; $ii)$ we compute the approximate solution with the retarded Green's function; $iii)$ we read out the $\beta$ coefficient by projecting out the negative frequency component.

\paragraph{Conformally coupled scalar.}~\\
In the presence of metric inhomogeneities (\ref{eq:metric}) the equation of motion to first order in perturbations is given by
\begin{equation}
\Box \varphi= -\partial^\mu(h_{\mu\nu} \partial^\nu \varphi) +\frac 1 2 \partial^\mu(h\eta_{\mu\nu} \partial^\nu \varphi)-\frac {\varphi} {6}(\partial^{\mu} \partial^{\nu} - \eta^{\mu\nu} \Box) h_{\mu\nu}\,,
\end{equation}
As in the previous section this equation can be easily solved using the retarded Green function of flat
space. The Bogoliubov coefficients are then obtained from the coefficients of positive and negative frequency waves.

A quicker way to determine the Bogoliubov coefficients is to solve the equation in Fourier space,
\be
\begin{split}
&(-k_0^2 + \vec k^2) \varphi(k_0,\vec k)=J(k_0,\vec k) \\
&J(k_0,\vec k)\equiv  \int \frac{d^4q}{(2\pi)^4}\bigg[k^\mu \omega^\nu -\frac 1 2 \eta^{\mu\nu} k\cdot \omega+\frac{1}{6}( q^\mu q^\nu -q^2 \eta^{\mu\nu})\bigg]h_{\mu\nu}(q_0,\vec q)\,\varphi(\omega_0,\vec\omega)\,,
\end{split}
\ee
where $k^\mu=q^\mu+\omega^\mu$, here and in the following.
The Fourier transform of the unperturbed solution reads,
\be
\varphi^{\rm in}( \omega_0,\vec \omega)=\frac{2\pi}{\sqrt{2\omega}}\bigg[\delta(\omega_0- \omega)a_{\vec \omega}+\delta(\omega_0+ \omega)a^\dag_{-\vec \omega}\bigg]\,.
\ee
To first order in the perturbations the solution for field is then,
\be
\varphi^{\rm out}(k_0,\vec k)=\varphi^{\rm in}(k_0,\vec k) + \frac{ J(k_0,\vec k)|_{\varphi=\varphi^{\rm in}}
}{-(k_0+i\epsilon)^2+|\vec k|^2}\,.
\ee
where we have shifted the poles of the propagator in order to select retarded boundary conditions.

Let us note that this equation is at the operator level. To extract the Bogoliubov transformation we could perform the Fourier transform $k_0\to \tau$ and match to the coefficient of the positive frequency component. This is equivalent to taking the residue of the function at $k_0=|\vec k|$.
Selecting the residue can be interpreted as taking the final particle on shell so that $k^\mu k_\mu=0$ and similarly
the initial state has $\omega_\mu \omega^\mu=0$. This fact will apply to all the cases we are going to present.

Thus we introduce the on shell momenta
\begin{equation}
\label{eq:4vectors}
\tilde k^{\mu}=\begin{pmatrix}
k \\
\vec{k}
\end{pmatrix}\,,\quad \quad
\tilde \omega^{\mu}=\begin{pmatrix}
-\sqrt{k^2+q^2-2 k q \cos\theta} \\
\vec{k}-\vec{q}
\end{pmatrix}\,, \quad \quad
\tilde q^{\mu}=\begin{pmatrix}
k +\sqrt{k^2+q^2-2 k q \cos\theta} \\
\vec{q}
\end{pmatrix}\,,
\end{equation}
where $\theta$ is the angle between $\vec k$ and $\vec q$. Note that $q_\mu$ is determined by momentum conservation and satisfies $\tilde q_\mu \tilde q^\mu\ge 0$. We stick to this notation in the rest of the paper when we compute a Bogoliubov coefficient for massless particle production.

Using on shell momenta the Bogoliubov coefficient is simply
\be\label{eq:bogoscalar}
\beta^{0*}_{\vec k\vec \omega}= \frac{h_{\mu\nu}(\tilde q)}{2\sqrt{k \omega}}\bigg[\tilde k^\mu \tilde \omega^\nu +\frac{1}{12}(2  \tilde q^\mu  \tilde q^\nu +  \tilde q^2 \eta_{\mu\nu})\bigg]\,,
\ee
and similarly for $\alpha_{\vec k\vec \omega}$ that however we will not need.

\paragraph{Massless fermions.}~\\
Let us now repeat the computation for Weyl fermions. The equation of motion in curved space is just $\bar\sigma^a e^\mu_a (\partial_\mu +\omega_\mu)\psi=0$. As shown in appendix \ref{app:tensors} to first order in perturbations the equation of motion reads,
\be
i \bar\sigma^\mu \partial_\mu \psi =  \frac{i}{16} \bar\sigma^\mu (\sigma^\rho\bar\sigma^\sigma - \sigma^\sigma \bar\sigma^\rho) \big[ \partial_\rho h_{\mu \sigma}-\partial_\sigma h_{\mu \rho}\big] \psi + \frac i 2 h^{\mu\nu}\bar\sigma_\nu \partial_\mu\psi
\ee
that in Fourier space gives,
\be
\begin{split}
\bar\sigma^\mu k_\mu  \psi(k_0,\vec k)&=J(k_0,\vec k) \nonumber \\
J(k_0,\vec k)&\equiv - \int \frac{d^4q}{(2\pi)^4}\bigg[\frac{1}{16} \bar\sigma^\mu (\sigma^\rho\bar\sigma^\sigma - \sigma^\sigma \bar\sigma^\rho) (q_\rho \eta_\sigma^\nu -q_\sigma \eta^\nu_\rho) + \frac12 \bar\sigma^\nu \omega^\mu \bigg]h_{\mu\nu}(q_0,\vec q )\psi( \omega_0,\vec \omega)
\end{split}
\ee
Using the unperturbed flat space field solution,
\be
\psi^{\rm in}(\vec \omega, \omega_0)=2\pi \big[\xi_{\vec \omega}^{-}\delta(\omega_0-|\omega|)a_{\vec \omega}+\xi_{-\vec \omega}^{+}\delta(\omega_0+|\omega|)b^\dag_{-\vec \omega}\big]\,
\ee
We obtain the solution to first order in perturbations,
\be
\psi^{\rm out}(\vec k, k_0)=\psi^{\rm in}(\vec k, k_0) +\frac{ \sigma^\mu k_\mu J(k_0,\vec k)|_{\psi=\psi^{\rm in}}}{(k_0+i\epsilon)^2-|\vec k|^2}
\ee
As for the scalar the Bogoliubov coefficients can be extracted from the residue at $k_0=k$. One finds,
\be
\beta^{-*}_{\vec k\vec \omega}=h_{\mu\nu}(\tilde q)\,\, \xi_{\vec k}^{-*}\frac{\sigma^\alpha \tilde k_\alpha}{2k}\bigg[\frac{1}{16} \bar\sigma^\mu (\sigma^\rho\bar\sigma^\sigma - \sigma^\sigma \bar\sigma^\rho) (\tilde q_\rho \eta_\sigma^\nu -\tilde q_\sigma \eta^\nu_\rho) +\frac12 \bar\sigma^\nu \tilde\omega^\mu \bigg] \xi_{-\vec\omega}^+
\label{eq:bogofermion}
\ee
This describes the production of the fermion of negative helicity. 
For the negative helicity state the result is identical and can be obtained considering the conjugate equation of for $\bar \psi$.
Note that the production of particles preserves helicity. This is manifest for massless fermions due to invariance of the action under chiral rotations.

\bigskip\medskip

\paragraph{Gauge fields.}~\\
Let us now turn to the case of massless spin-1 fields, see also \cite{Maroto1}. 
 Here we proceed by going to the Coulomb gauge where $A_0=\nabla\cdot \vec A=0$. The equation of motions with metric perturbations is then
\begin{equation}
\Box A^\nu=\partial^\alpha(h_{\mu\alpha} F^{\mu\nu}) - \partial_\alpha(h_\mu^\nu F^{\mu\alpha})-\frac 1 2 \partial_\mu (h F^{\mu\nu})\,.
\end{equation}
so that in Fourier space,
\be
(-k_0^2+k^2)A^\nu(k_0,\vec k)=J^\nu(k_0,\vec k)
\ee
where the source can written as,
\be
\begin{split}
J^\nu(k_0,\vec k)&\equiv - \int \frac{d^4q}{(2\pi)^4}Z^{\alpha\beta\sigma\nu}(k,q) h_{\alpha\beta}(q_0,\vec q)\,A_\sigma(\omega_0,\vec\omega)\nonumber \\
Z^{\alpha\beta\sigma\nu}&= \omega^\alpha k^\beta \eta^{\sigma\nu} - k^\beta \omega^\nu \eta^{\alpha\sigma} - \eta^{\alpha\nu}\omega^\beta k^\sigma +\eta^{\alpha\nu}\eta^{\beta\sigma}k\cdot \omega -\frac12 \eta^{\alpha\beta}\eta^{\sigma\nu}k\cdot \omega +\frac12 \eta^{\alpha\beta}k^\sigma \omega^\nu\,.
\end{split}
\ee
The construction of the solution at first order in the metric fluctuations goes as before. The unperturbed solutions is given by
\be
A^{{\rm in} \mu}(\omega_0,\vec \omega) = \sum_{\lambda=1,2}\frac{2\pi}{\sqrt{2\omega}}\big[\varepsilon^\mu_{\vec \omega\,\lambda}\delta(\omega_0-\omega)a_{\vec \omega,\lambda}+\varepsilon^{\mu*}_{-\vec \omega\,\lambda}\delta(\omega_0+\omega)a^\dag_{-\vec \omega,\lambda}\big]\,.
\ee
The two polarization vectors are orthogonal to each other, they satisfy $\omega_\mu \varepsilon^\mu_\lambda(\omega)=0$ and they are such that (in this gauge) $\varepsilon^0=0$. The solution for the field to first order is then,
\be
A^{{\rm out},\nu}(k_0,\vec k)=A^{{\rm in}\nu}(\vec k, k_0) + \frac{J^\nu(k_0,\vec k)|_{A=A^{\rm in}}}{-(k_0+i\epsilon)^2+|\vec k|^2}\,.
\ee
This is totally analogous to the scalar case. As in that case we compute the Bogoliubov coefficients by projecting onto the out-basis $\bar a_{\vec k, \lambda}$, which just amount to take the residue of the new term at $k_0=|\vec k|$. While any basis is equivalent the formulae become particularly transparent choosing the helicity basis
where the polarizations are eigenstates of spin $\vec k \cdot \vec J$.\footnote{Polarizations are constructed as eigenstates of the operator $\sim\vec k\cdot \vec J/k$, with eigenvalues $\lambda=\pm J$, see appendix \ref{app:tensors}.} Using this basis the only non zero coefficients are the ones that preserve helicity. 
This identical to the fermion case where however the result follows automatically from chiral invariance. We find,
\be
\beta^{\pm *}_{\vec k\vec \omega}=\frac{h_{\mu\nu}(\tilde q)}{2\sqrt{k\omega}}\, \varepsilon^{
\rho*}_{\vec k, \pm}Z^{\mu\nu}_{\sigma\rho} \varepsilon^{\sigma*}_{-\vec \omega,\pm}
\label{eq:bogogauge}
\ee
where as usual the momenta are evaluated on-shell.

\paragraph{Summary.}~\\
From the above explicit case, we can summarize our findings in a very clear formula. By inspection, we notice that the Bogoliubov coefficients for the production of a particle with helicity $|\lambda|=J$ have the very simple form
\be\label{eq:bogo-exclusive}
\beta_{\vec{k} \vec{\omega}}^{\lambda} = \frac{h_{\mu\nu}(\tilde q)}{2\sqrt{k\omega}}\times A_J^{\mu\nu}\,.
\ee

In this form, the symmetric tensor $A_J^{\mu\nu}$ has to be computed with massless on-shell momenta $\tilde k$ and $\tilde \omega$ ($\tilde \omega_0<0$) so that $\tilde q^\mu \tilde q_\mu>0$.  
The effect of the perturbation is factorized, so that one can compute the structure $A_J$ once and for all. Let us note that in this form the Bogoliubov coefficients have just the form of an on-shelll amplitude.
In fact an alternative way to derive the results is compute the Feynman amplitude for the creation of particle pairs in the external field background as in \cite{Campos:1991ff,Cespedes:1989kh}.
More details will appear elsewhere.

In the massless limit it is convenient to use the helicity basis for $\lambda$, where all the polarization vectors spinors are eigenstate of the helicity operator. In this basis one can show that $A_J$ vanishes when helicity violating amplitudes are computed. The cancellation is manifest upon symmetrization of the tensor $A_J^{\mu\nu}$ in the Lorentz indices, since it is anyway contracted with the symmetric $h_{\mu\nu}$. 
These structures are diagonal in the helicity basis but depend on spin. 
We ascribe this to helicity conservation, which holds for any metric background $h_{\mu\nu}$. Interestingly, we just have very few expressions. We also notice, without surprise, that $\eta_{\mu\nu}A^{\mu\nu}_J=0$.

\bigskip

Eventually, we can pack together all our derivation for the case of conformally coupled matter in presence of fluctuations. Using eq.~\eqref{eq:Ndiff0}, we have our master formula for Bogoliubov production from inhomogeneities
\begin{tcolorbox}[colback=lightergray, colframe=white, boxrule=0pt, left=0mm, right=0mm]
\be\label{eq:master-excl}
\frac {dN_\lambda}{dk d(\cos \theta_k) d\phi_k}=  \frac {k^2}{(2\pi)^3} \int \frac{d^3 q}{(2\pi)^3}h_{\mu\nu}(\vec q, k+\omega) h_{\rho\sigma}^*(\vec q,  k+\omega) \times \frac{A_J^{\mu\nu}A_J^{\rho\sigma\, *}}{4 k \omega}
\ee
\end{tcolorbox}
~\\
This formula is more exclusive than the result of the 1PI effective action -  the other grey box in our paper -  of eq.~\eqref{eq:Nparticles}. A price to pay for this exclusiveness is the fact that explicit calculation can be carried out for weakly coupled conformally coupled matter, while the 1P1 effective action encompasses also strongly coupled CFTs. It has however some strong similarities with that, some of which are rather obvious. In the following subsection we comment on how to show they are equivalent.

\subsection{Equivalence with effective action computation}

The  differential result computed for a classical metric perturbation looks on the surface very different from the result 
obtained in section \ref{sec:inout}.
To connect with the effective action computation we need to consider the inclusive number of particles produce integrating the phase space $d^3k$ of the particles produced.
The Bogoliubov coefficients depend on $h_{\mu\nu}(q\,,q_0 )$ that suggests to interpret the argument of the time Fourier transform as the energy $q_0$ in the 1PI effective action (\ref{eq:G1PI}).
We thus introduce the change of variables, 
\begin{equation}
q_0= k+\omega\equiv k +\sqrt{k^2+q^2 -2 q k \cos\theta} 
\label{eq:changevariables}
\end{equation}
so that,
\begin{equation}
k(q_0)= \frac {q_0^2-q^2}{2(q_0-q \cos\theta)}\,,~~~~~~~~\frac {dk}{dq_0}=\frac {q^2+q_0^2-2 q q_0 \cos \theta}{2(q_0-q \cos\theta)^2}
\end{equation}
where $k>0$ implies $q_0> q$. The on-shell momenta in (\ref{eq:4vectors}) become,
\begin{equation}
\tilde{k}^{\mu}=\frac{q_0^2 - q^2}{2 (q_0 - q \cos\theta)}\begin{pmatrix}
 1\\
n_{\vec{k}}
\end{pmatrix}\,,\quad \quad
\tilde \omega^{\mu}=\begin{pmatrix}
-|\vec k -\vec q| \\
\vec{k}-\vec{q}\,,
\end{pmatrix}
\quad \quad
\tilde q^{\mu}=\begin{pmatrix}
q_0\\
\vec{q}
\end{pmatrix}
\end{equation}
With this change of variables the integral with respect to $\theta$ can be performed analytically. 
One can explicitly check using (\ref{eq:bogoscalar}),(\ref{eq:bogofermion}),(\ref{eq:bogogauge}) and performing the angular integrals that,
\begin{equation}
\frac {dN}{d q_0}= \frac {q_0^2}{4\pi^2} \int d(\cos \theta_k) \int \frac{d^3q}{(2\pi)^3} |\beta_{\vec k\vec \omega}|^2 
\label{eq:generalBogo}
\end{equation}
is equivalent to eqs. (\ref{eq:Nparticles}),(\ref{eq:Nparticles2}).
While the explicit computation is non trivial the equivalence between the two computation can be proven on general grounds.  
We provide a derivation in the presence of inhomogeneities in the appendix.

Let us note the computation above clarifies the meaning of $q_0$: this is the energy of the pair of particles produced by the decay of the perturbation $\vec{q}$. 
The only actual difference is that the computations using the Bogoliubov transformations allows
to also compute the  distribution of single particles. For example since $q_0>q$ the energy of the particles pair is bounded by the perturbation
while distribution of 1 particle extends up to $k=0$. The differential distribution is not accessible to the effective action computation
because this really depends on the 3-point function of the energy momentum tensor that contrary to the 2-point function 
is not fixed by conformal invariance \cite{Osborn:1993cr}. As a consequence the computation with Bogoliubov coefficients must be done 
case by case with intermediate results that appear very different. It is interesting that the effective action computation
allows to determine the production of sectors without a lagrangian formulation where the Bogoliubov coefficients could not be determined.

\subsection{Scalar and tensor perturbations}\label{sec:scalar-tensor-general}

Eq.~\eqref{eq:master-excl} holds for general metric perturbations that vanish in the far past and far future. 
In what follows we will be mostly interested in scalar and gravity wave perturbations, potentially present 
during the cosmological evolution of our Universe. Such fluctuations can be parametrized as follows
\be\label{eq:conformal-gauge}
ds^2 = a^2(\tau) [1+2\Phi(\tau,\vec x)]d\tau^2 - a^2(\tau)  [\delta_{ij}(1-2\Psi(\tau, \vec x))-h_{ij}(\tau, \vec x)]dx^i dx^j\,.
\ee
Here we have chosen conformal Newtonian gauge for scalar perturbations while tensor perturbations are transverse traceless $\partial^i h_{ij}=h^i_i=0$.
The scalar perturbations are encoded in the two functions $\Psi$ and $\Phi$. 
Notice also that in absence of anisotropic stress-energy tensor components $\Phi=\Psi$, but we will not need to make this assumption. 
The interaction lagrangian  in this gauge reads
\be
\Delta \mathscr{L}_{\rm int}= \frac{\Psi+\Phi}{2} \delta_{\mu\nu} T^{\mu\nu} +  \frac{\Phi-\Psi}{2} \eta_{\mu\nu} T^{\mu\nu} + \frac{h_{ij}}{2}T^{ij}\,,
\ee
where $T^{\mu\nu}=2/\sqrt{-g} \delta S/\delta g_{\mu\nu}$ is the stress-energy tensor of the CFT coupled to gravity. In the above equation we see that only specific linear combinations of the gravitational potential couple to the matter sector. It is then convenient to define the variables\footnote{In other contexts the scalar $\Sigma$ is known as the dilaton.} 
\be
\Theta(\tau,\vec x)\equiv\Phi + \Psi\,,\quad\quad \Sigma(\tau,\vec x)\equiv\Phi - \Psi\,.
\ee
For conformally coupled matter, Weyl invariance implies $\eta_{\mu\nu}T^{\mu\nu}=T=0$ as an operator statement. As a consequence the perturbation $\Sigma$ can be removed 
through a Weyl transformation of the fields at the classical level, if the action is invariant under Weyl transformations. 
We can thus neglect the term proportional to $\Sigma$ that does not lead to the production of massless conformally coupled particles. This can also be checked explicitly using  the formulas 
derived in the previous section because the Weyl tensor vanishes for $\Sigma$. 

The tensor perturbations are transverse traceless but not necessarily on shell to account for their production and cosmological evolution.
The general form of tensor perturbations can be chosen as,
\be
h_{ij}(\tau,\vec x) = \sum_{\pm} \int \frac{d^3q}{(2\pi)^3}e^{ i \vec q \cdot \vec x} h^{\pm}(\tau, \vec q) \epsilon_{ij}^{\pm}(\vec q)+h.c.\,.
\label{eq:GWB}
\ee
where $\epsilon_{ij}^\pm$ are the two symmetric transverse traceless polarization matrices for a gravity waves of wavenumber $\vec q$.
In what follow we will use the helicity basis for the gravitational field, see  \cite{Weinberg:2008zzc,kopp2}.
Explicit expression for the polarizations are reported in appendix \ref{appC}.

\bigskip

With the ingredients above we can proceed with the computation of the Bogoliubov coefficients. It just amounts to evaluate eq.~\eqref{eq:bogo-exclusive} for the choice of perturbations in the equation above. 
Denoting with $|\lambda|=J$ is the spin of the particle the result can be written as,
\be
\beta^\lambda_{\vec k\vec \omega}=  \frac{\Theta(k+\omega,\vec q)}{2\sqrt{k\omega}} (B_J)_{\vec k\vec \omega}  \,\, + \frac{h^{+}(k+\omega,\vec q)}{2\sqrt{k\omega}} (C_J^+)_{\vec k\vec \omega}+ \frac{h^{-}(k+\omega,\vec q)}{2\sqrt{k\omega}} (C_J^{-})_{\vec k\vec \omega}\,.
\label{eq:amp}
\ee
Helicity conservation implies that only helicity preserving processes are allowed. Moreover processes with opposite helicities (including the one of the gravitational field) have the same Bogoliubov coefficients (up to phases) due to parity.  We collect the results in the Table \ref{tab:amp}, that can be taken as the main result for any cosmological application where scalar and tensor fluctuations are generated (from inflation or phase transitions, for example). As manifest from \eqref{eq:bogo-exclusive}, the expressions are just function of $\vec k$ and $\vec \omega=\vec k -\vec q$.

\renewcommand{\arraystretch}{1.2}
\begin{table}[h!]
\centering
\setlength{\tabcolsep}{2.5pt} 
\resizebox{\textwidth}{!}{ 
\begin{tabular}{c|c|c}
\multicolumn{3}{>{\columncolor{lightergray}}c}{$J=0$}\\ \hline
$B_{0}$ & $C_{0}^{+}$ & $C_{0}^{-}$\\ \hline
$ \displaystyle \frac{3(k-\omega)^2-q^2}{6}$ & $ \displaystyle \frac{4k^2 q^2- \left(k^2+q^2-\omega ^2\right)^2}{4\sqrt{2} q^2}$ & $\displaystyle \frac{4k^2 q^2- \left(k^2+q^2-\omega ^2\right)^2}{4\sqrt{2} q^2}$ \\ \hline
\multicolumn{3}{>{\columncolor{lightergray}}c}{$J=1/2$}\\ \hline
$B_{1/2}$ & $C_{1/2}^{+}$ & $C_{1/2}^{-}$\\ \hline
$\displaystyle  \frac12 (\omega-k)\sqrt{q^2-(k-\omega)^2}$ & 
$\displaystyle  \frac{((k+\omega)^2-q^2)(k-q-\omega)\sqrt{q^2-(k-\omega)^2}}{4\sqrt{2}q^2}$ &  
$\displaystyle  \frac{((k+\omega)^2-q^2)(k+q-\omega)\sqrt{q^2-(k-\omega)^2}}{4\sqrt{2}q^2}$ \\ \hline
\multicolumn{3}{>{\columncolor{lightergray}}c}{$J=1$}\\ \hline
$B_{1}$ & $C_{1}^{+}$ & $C_{1}^{-}$\\ \hline
$\displaystyle \frac12 (q^2-(k-\omega)^2)$ & 
$\displaystyle  \frac{(k+q-\omega)^2((k+\omega)^2-q^2)}{4\sqrt{2}q^2}$ & 
$\displaystyle  \frac{(-k+q+\omega)^2((k+\omega)^2-q^2)}{4\sqrt{2}q^2}$ \\ 
\end{tabular}
}
\caption{Bogoliubov coefficients in \eqref{eq:amp} for different spins produced by scalar and tensor perturbations (up to irrelevant phases). 
In the latter case, the result depends on the polarization. For spin-1/2 and spin-1 processes with opposite helicities gives the same result.}
\label{tab:amp}
\end{table}

Table \ref{tab:amp} contains all the information needed for the calculation for scalar and tensor perturbations. Notice that so far we have made no assumptions on the nature of the perturbation $h_{\mu\nu}$, its statistical properties for example, made exception for the fact that it disappears in the far past and far future.

\subsection{Relation to previous work}

Particle production from inhomogeneities was studied in several papers with different methods and different goals. 
To the best of our knowledge all these works missed the relevance of the two-point function of the energy momentum 
tensor that is manifest using the Schwinger 1PI effective action approach. In the massless limit this actually only depends 
on the central charge of the CFT regardless of its interactions. 

Particle production from cosmological perturbations were first considered by Starobinski and Zeldovic
in \cite{Zeldovich:1977vgo} first noticing the connection with the Weyl tensor. They considered a conformally coupled massless scalar in a Bianchi type-I described by
\begin{equation}
ds^2= a^2(\tau) [d\tau^2 - (1+h_x(\tau)) dx^2- (1+h_y(\tau)) dy^2 -(1+h_z(\tau)) dz^2]\,,~~~~~~~~\sum_{i=x,y,z} h_i(\tau)=0\,,
\end{equation}
and computed the Bogoliubov transformation induced by the background for a conformally coupled field.
In general their result can be written in the following form,
\begin{equation}
na^3= \frac{c_J}{7680 \pi}\int d\tau \sum (h_i'')^2= \frac{c_J}{3840 \pi} \int d\tau W_{\mu\nu\rho\sigma}^2\,.
\end{equation}
In this case, since the perturbation is  space independent, the Bogoliubov transformation does not mix modes of different momentum, 
similarly to mass perturbations. Our formulas reproduce this result in the limit $q=0$ for gravity waves. In this limit the kernel is proportional to $k^2$ 
so that the integral in $k$ produces $\delta''''(\tau-\tau')$. Integrating by parts the result above is reproduced.

In a series of papers Hu and collaborators studied particle production from inhomogeneities, see \cite{Hu:2020luk} for a review of the results.
In \cite{Calzetta:1986ey} the abundance of particles was computed using the Schwinger-Keldish formalism (also known as in-in formalism) to
compute the effective action that sources the Einstein equations. The use of in-in formalism is motivated by the fact that in cosmology 
only the incoming states can be defined. For conformally coupled fields however we can remove the expansion of the universe from the 
equation of motion so that the problem reduces to a flat space problem. This allows the use of the simpler in-out formalism that in particular 
allow to compute particle production through the 1PI effective action. Moreover the energy distribution requires further work.

In \cite{Campos:1991ff,Cespedes:1989kh} (see also \cite{Parker_Toms_2009} and references therein) particle production was derived also from an S-matrix point of view as the decay of the perturbation to 2 particles. 
Our results written in terms  of amplitudes (\ref{eq:amp}) is similar in nature  to that work, although we have a taken a different route in deriving the particle number.

\bigskip

We leave to the future the extension of our formalism to massive particles. We expect however this mechanism of production to be mostly effective
when the particle masses are negligible. Since the production is suppressed by the square of the amplitude of the perturbation particle production 
due to the mass, that produces ${\cal O}(1)$ non-adiabaticity is likely to dominate. 

\section{Stochastic backgrounds}
\label{sec:stochastic}

Having determined general formulas for particle production from inhomogeneities it is interesting to study the effect of a stochastic background, where formulas greatly simplify. We emphasize that our observables depends on un-equal time correlators. For a stochastic  homogeneous and isotropic background we have
\begin{equation}
\langle h_{\mu\nu}(\vec q,q_0) h_{\rho\sigma}(\vec q\,',q_0') \rangle= (2\pi)^3 \delta^3(\vec q+\vec q\,'\,)P_{\mu\nu\rho\sigma}(q\,,q_0\,,q_0')\,.
\end{equation}
Taking the average of eq. (\ref{eq:Nparticles}) we thus find,
\begin{equation}
n\equiv \frac N V= \frac {c_J}{15360\pi}\int \frac {d^4 q}{(2\pi)^4}\theta(q_0^2-q^2) P_{\mu\nu\rho\sigma}(q\,,q_0\,,-q_0) \Pi^{\mu\nu\rho\sigma}(q\,,q_0\,,-q_0)
\label{eq:Nstochastic}
\end{equation}
where we used $(2 \pi)^3 \delta^3(0)\equiv V$. Or, more exclusively, we can take eq.~\eqref{eq:master-excl} and get
\be
\frac {dn_\lambda}{dk d(\cos \theta_k) d\phi_k}=  \frac {k^2}{(2\pi)^3} \int d^3 q P_{\mu\nu\rho\sigma}(q\,,q_0\,,-q_0) \times \frac{(A_J)^{\mu\nu}(A_J)^{\rho\sigma\, *}}{4k \omega}
\ee
As expected for a stochastic background the quantity that is well defined 
is the number density because particles are produced everywhere.
Note that under the assumption of isotropy and homogeneity the angular dependence is trivial so that the differential formulas 
contain no extra information in this case.

We now provide explicit formulas for scalar and tensorial perturbations.

\subsection{Scalar perturbations}

As discussed in section \ref{sec:scalar-tensor-general}, particle production of conformal matter only depends on the gravitational potential $\Theta=\Phi+\Psi$, the orthogonal combination being
Weyl flat. We assume the power spectrum
\begin{equation}
\langle \Theta(\vec q\,,q_0) \Theta(\vec q\,'\,,-q_0) \rangle= (2\pi)^3 \delta^3(\vec q+\vec q\,') \frac {2\pi^2}{q^3}\Delta_\Theta(q\,,q_0) \,.
\end{equation}
Note again that this quantity depends on the un-equal time correlation function.

With our formalism, for a stochastic background of scalar or tensor perturbations the differential abundance of particles of spin-$J$ at late times is then given by,
\begin{equation}
\frac{d (n_J a^3)}{d\log k}= \frac {k^3}{4\pi^2} \int d(\log q)\int d(\cos \theta)\Delta_{\Theta}(q, k+\omega) K_J^\Theta[k, q, \cos \theta]\,.
\ee
This formula is nicely factorized into the power spectrum that depends on cosmology, and a kinematical function that depends upon the particle spin. 
We have therefore introduced a Kernel function $K_J^\Theta$ that depends on three variables. They can be chosen as $|\vec k|$, $|\vec q|$, $\cos \theta$ (alternatively we can replace $\cos\theta=(k^2+q^2-\omega^2)/(2qk)$ with the constraint $|k-q|<\omega< k+q$). 

For scalar perturbations $\Theta$, the kernels for different spin are given by
\be
\begin{split}
K_0^\Theta& = \frac{|B_0|^2}{4 k\omega} \stackrel{q\to 0}{=}  \frac{q^4 (3 \cos (2 \theta )+1)^2}{576 k^2}+\dots\\
K_{1/2}^\Theta &=  2\frac{|B_{1/2}|^2}{4 k\omega}\stackrel{q\to 0}{=} \frac{q^4 \sin^2(2\theta)}{32k^2}  +\dots \\
K_1^\Theta &= 2 \frac{|B_1|^2}{4 k \omega}\stackrel{q\to 0}{=} \frac{q^4 \sin ^4(\theta )}{8k^2}+\dots
\end{split}
\label{eq:kernelscalar}
\ee
These are just the square of the amplitude of each individual processes (\ref{eq:amp}). A factor of two is included to account for particles with helicity. 
Scalar kernels scale a $q^4/k^2$ so that particle production vanishes in the limit $q \to 0$. This behaviour follows from the fact that
a spatially constant scalar perturbation corresponds to a Weyl flat metric.  By performing the integral in $\cos\theta$ (in the limit $q\to 0$), 
one can show explicitly that the Kernels are normalized to the central charges of the corresponding field. 

In fact this result holds not just for $q/k\ll1$. Indeed from eq. (\ref{eq:Nstochastic}) we obtain the number density of particles,
\begin{tcolorbox}[colback=lightergray, colframe=white, boxrule=0pt, left=0mm, right=0mm]
\begin{equation}\label{eq:scalar-stoc}
\frac {d(n_J a^3)}{d q_0}= \frac {c_J}{960\pi^2}  \int d(\log q)   \Delta_\Theta(q\,,q_0) q^4\,.
\end{equation}
\end{tcolorbox}~\\
Here $q_0$ is the sum of the energy of particles produced.
Using the change of variables in eq.~(\ref{eq:changevariables}) and performing the angular integral one can check that eq.~(\ref{eq:kernelscalar}) reduces to the equation above.
Therefore the abundance is completely determined by the central charge and power spectrum.

\subsection{Tensor perturbations}

Let us now consider a gravity wave  background parametrized by (\ref{eq:GWB}). For a stochastic (unpolarized) gravity wave background,
\begin{equation}
\langle h^\pm(\vec{q}\,, q_0) h^{\pm\,*}(\vec{q}\,'\,, -q_0)\rangle= (2\pi)^3 \delta(\vec{q}-\vec q\,') \frac {2\pi^2}{q^3}\Delta_h(q\,, q_0)\,
\end{equation}

The abundance of particles produced is now given by,
\be
\frac{d (n_Ja^3)}{d\log k}= \frac {k^3}{4\pi^2} \int d(\log q)\int d(\cos \theta)\Delta_{h}(q, k+\omega) K_J^h[k\, q, \omega]\,.
\end{equation}
The structure is similar to the case of scalar perturbations,
and the kernels are now given by
\begin{eqnarray}\label{eq:tensor-kernels}
&&K_0^h = \frac {|C_0^+|^2+|C_0^-|^2}{4 k \omega}\stackrel{q\to 0}{=} \frac{k^2}{4} \sin ^4\theta +\cdots \nonumber \\
&&K_{1/2} ^h=2\frac {|C_{1/2}^+|^2+|C_{1/2}^-|^2}{4 k \omega}\stackrel{q\to 0}{=} \frac{k^2}{2}(1-\cos^4\theta)+\cdots 
\nonumber \\
&& K_1^h =2 \frac {|C_1^+|^2+|C_1^-|^2}{4 k \omega}\stackrel{q\to 0}{=} \frac{k^2}{2}(1+6\cos^2\theta +\cos^4\theta))+\cdots
\end{eqnarray}
Notice that differently from scalars perturbations the kernels grow in this case as $k^2$ while for scalar perturbations they go as $q^4/k^2$. This  behaviour 
is at the origin of the different ultraviolet sensitivity of scalar and tensor productions. Notice again that fermions and vectors have a factor of two coming from the two helicities. 

For the total number of particles we find in this case
\begin{tcolorbox}[colback=lightergray, colframe=white, boxrule=0pt, left=0mm, right=0mm]
\begin{equation}\label{eq:total-tensor}
\frac {d(n_Ja^3)}{dq_0}= \frac{c_J}{640\pi^2}  \int  d(\log q) \theta(q_0^2-q^2)  (q_0^2-q^2)^2 \Delta_h(q\,, q_0)\,.
\end{equation}
\end{tcolorbox}
Let us comment on the UV dependence of the integrals. For scalar perturbations the integral over $q_0$ is finite if the power spectrum goes to zero
faster than $1/q_0$. We will see in explicit examples that this is typically the case. For the gravity waves the kernel goes as $q_0^4$ leading to more UV sensitivity of the 
integral. Because of this in \cite{kopp2} it was proposed that on physical grounds the integral should be cut at $q_0\sim q$. We note however that particle production 
is just determined by the Weyl curvature. As a consequence in smooth background that asymptotically reduce to flat space the number of particles will always be finite
without the need of regularisation. We will show an explicit example for gravity waves produced during inflation.

\section{Metric perturbations from Inflation}
\label{sec:inflation}

The general formulae for the production from inhomogeneities depend on the power spectrum at unequal times.
The time dependence is model dependent being determined on how the signal is generated. For example during a phase transition inhomogeneities are sourced at a specific time and then evolve according to free equation of motion. Since particle production is controlled by non adiabaticity of the evolution particle production is then dominated at the time of phase transition. 

In this section we consider the case where perturbations are produced during inflation, extending the analysis in  \cite{Garani:2024isu}.
We consider single field inflationary scenarios  where  the two physical perturbations are the curvature scalar $\zeta(\tau\,,\vec x)$ and tensor perturbations $h_{ij}(\tau\,,\vec x)$
that correspond to gravity waves. During inflation perturbations are produced at horizon exit and roughly  remain constant until horizon re-entry during reheating or radiation domination.
We will focus on scalar perturbations that could lead to the production of the DM abundance.
They are described by the power spectrum at the end of inflation
$$
\langle \zeta_{\vec q} \zeta^*_{\vec q} \rangle= \frac{2\pi^2}{q^3}\Delta_\zeta (q)\,.
$$
Here $\zeta_{\vec q}$ is the Fourier transform evaluated at the end of inflation. Precise determinations of the CMB observables and large scale structure determine the amplitude and tilt of the power spectrum on 
 cosmological scales, $q^{-1}\approx \mathrm{Gpc-Mpc}$. One finds $A_\zeta(q_*)=2.1\times 10^{-9}$ at the pivot scale $q_*=0.05/$Mpc with a small tilt.
The amplitude could however be much larger at shorter scales where it is mostly unconstrained. Given that  inflation should last around 60
e-foldings this produces scalar perturbations with comoving momenta up to $q_{\rm max} \sim e^{60} H_0 \sim 10^{-7}$ eV.
Such mechanism allows for example to produce primordial black holes if the amplitude of the power spectrum is  enhanced as some scale, $\Delta(q_*)\sim 10^{-2}$.
In this case the mass of the black is controlled by $q_*$ so asteroidal mass black-holes that could constitute the totality of DM require
$q_*\sim 10^9\, {\rm Mpc}^{-1}$. The enhanced power spectrum would also produce particles. This allows to reproduce the DM abundance
especially if the power spectrum is enhanced towards the end of inflation.

\subsection{Scalar perturbations}

Our formulas depend on the scalar gravitational potential $\Theta\equiv \Phi+\Psi\approx 2\Psi$ where in the second step we assume negligible anisotropic stress.
We recall that a gauge invariant definition requires the introduction of matter fluctuations, as for example in conformal Newtonian gauge the gauge invariant combination is
\be
\zeta\equiv -\Psi - \frac{H}{\dot{\bar\rho}}\delta \rho\,.
\ee
The relation between $\zeta$ and the gravitational potential depends on the cosmology.
Working in conformal newtonian gauge one can show that $\Psi=0$ during inflation. 
This means that no particle production occurs during inflation. This is also expected because without 
time dependence no particles are produced.

The curvature perturbation $\zeta$ determine the initial condition of the perturbations in standard cosmology.
The previous values are matched on super-horizon scales to the values of $\Phi$ and $\Psi$ appearing in our eq.~\eqref{eq:conformal-gauge}. 
We do the matching on scales $q\tau\to 0$, and we read out the following conditions
$$
\Psi_{\vec q}|_{q\tau\to 0}=\Phi_{\vec q}|_{q\tau\to 0}=\kappa_\alpha \zeta_{\vec q}\,,
$$
where $\kappa_{\rm rad}=-2/3$ in radiation dominance and $\kappa_{\rm mat}=-3/5$ in matter dominance. 
On these scales we completely neglect the anisotropic stress, so that the initial evolution in standard cosmology starts out with $\Phi=\Psi$.
It is convenient then, to write $\Phi$ at all times in terms of the matching condition, as follows
\begin{equation}
\Phi(\vec q,\tau)=   T(q\,, \tau)\,\zeta_{\vec q}\,,
\end{equation}
where $T(q,\tau)$ is a transfer function that solves the cosmological evolution equation of the gravitational potential. 
During inflation $\tau <0$, the transfer function goes to zero. The transfer function for the gravitational potential $\Phi$, can be written -- when the universe is dominated by one fluid with sound speed $u_s^2$ -- in the following form
\begin{equation}
T''(q\,,\tau)+3 \frac {a'}{a}(1+u_s^2) T'(q,\tau) + u_s^2 q^2  T(q\,,\tau)=0\,,\quad T(q\,,0)=\kappa_\alpha\,,\quad T'(q\,,0)=0\,.
\label{eq:Phi}
\end{equation}
In more general phases, one needs to solve the Einstein equations to determine $T(q,\tau)$, and this can be done precisely with numerical codes. Let us notice that all the unknowns about the cosmological evolution of our universe are encoded in the transfer function, which only well measured at cosmologically large scales.
The explicit expression of $T(q, \tau)$ is all we need to compute $\Delta(\vec q, q_0)$ which is given by
\be
\Delta(\vec q, q_0) = \Delta_\zeta(\vec q\,)\times |\mathcal{I}(q\,,q_0)|^2\,,\quad \text{with} \quad\mathcal{I}(q\,,q_0)\equiv  \int_{-\infty}^\infty d\tau \,e^{-iq_0\tau}T(q,\tau)\,.
\ee
This is just proportional to the square of the time Fourier transform of the transfer function. The function $\mathcal{I}(q\,,q_0)$ depends on the cosmic evolution, such as reheating and standard cosmology, while the primordial power depends just on inflation. 

At this point we can arrive at a particularly simple expression for the number density using eq.~\eqref{eq:scalar-stoc}. By performing the integral in $q_0$ we have the following - extremely simple - final result
\begin{tcolorbox}[colback=lightergray, colframe=white, boxrule=0pt, left=0mm, right=0mm]
\begin{equation}
na^3 =c_J \frac  {A^{\zeta}}{4\pi^2}  \int d (\log q) q^3 \Delta_{\zeta}(q) 
\label{eq:abinflation}
\ee
\end{tcolorbox}
The overall factor $A^\zeta$ - a number - depends on the cosmology and it is given by
\begin{equation}
A^{\zeta}\equiv  \frac 1 {240}\int_1^{\infty} dx  |\mathcal{I}(1,x)|^2\,.
\label{eq:Azeta}
\end{equation}
We now discuss two different scenarios for the cosmological evolution. First, a case with instantaneous reheating, such as that the evolution of the modes is just experiencing standard cosmology. Second, a case with reheating of finite duration $\tau_R$. In our notation $\tau=0$ signals the end of inflation. We consider evolution prior to matter radiation equality, since we have in mind applications for DM. Extensions to all time scales are easy to get.

\begin{figure}
\begin{center}
\includegraphics[width=0.48\textwidth]{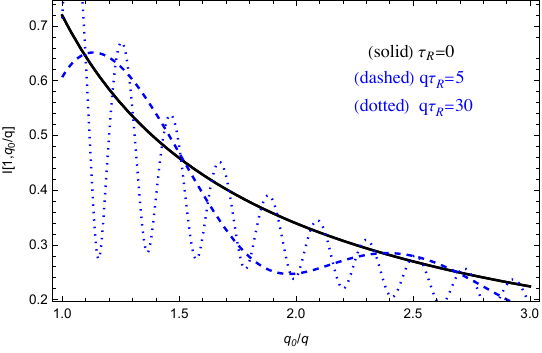}\quad 
\includegraphics[width=0.48\textwidth]{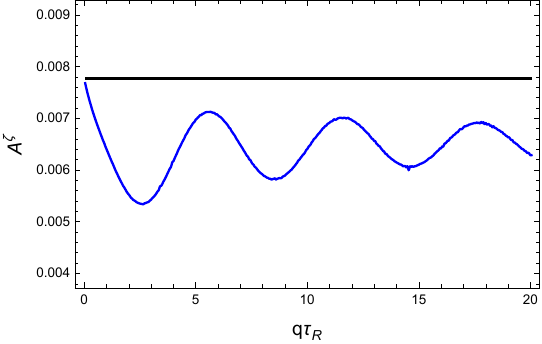}\quad 
\\
\caption{\label{fig:transfer} \textit {Left: Scalar transfer function for modes that re-enter the horizon during matter and  radiation domination (blue dashed $q \tau_R=5$,
blue dotted $q \tau_R=30$).
Right: Plot of $A^{\zeta}$ as function of reheating time for modes that re-enter the horizon during matter domination (blue) or radiation domination (black).
We fix $q=1$.}}
\end{center}
\end{figure}

\paragraph{Instantaneous reheating.}~\\
Let assume that at $\tau=0=\tau_R$ the universe starts a phase of radiation.
In this case $\tau_R=0$, and radiation dominance starts right at the end of inflation. During radiation the transfer function is simplify given by
\begin{equation}
T_{\rm rad}(q\,, \tau)=  2  \frac{\tau  q u_s \cos (\tau  q u_s)-\sin (\tau  q u_s)}{\tau ^3 q^3 u_s^3}\,,\quad u_s=1/\sqrt{3}\,.
\end{equation}
This can be Fourier transform to get the function $\mathcal{I}_{\rm rad}(q\,,q_0)$
\begin{equation}
\mathcal{I}_{\rm rad}(q\,,q_0)=i\frac{  q_0 q u_s -(q_0^2- u_s^2 q^2) \tanh ^{-1}\left(\frac{q u_s}{q_0}\right)}{ q^3 u_s^3} 
\end{equation}
This function behaves as  $\mathcal{I}_{\rm rad}(q\,,q_0)= i \frac 2 {3 q_0} +\dots$ when $q_0\gg q$, and this is enough to make the integral in eq.~\ref{eq:Azeta}) convergent. For modes that re-enters the Hubble radius in radiation dominance, we find
\begin{equation}
A^{\zeta}_{\rm rad}=0.008\,.
\end{equation}

\paragraph*{Finite duration of reheating.}~\\
The discussion above strictly applies to modes that re-enter the horizon during radiation domination. 
However -- as we will now show -- the result is almost unchanged for the modes that re-enter the horizon during reheating. 
This is particularly relevant because the abundance is likely dominated by the largest momenta.

In order to demonstrate this we consider a finite duration of reheating, approximated by a phase of matter dominance, that extends from the end of inflation at time $\tau=0$ to time $\tau_R$, where the standard radiation phase begins. Approximating the  reheating as a phase of matter domination, implies that $\Phi$ is constant throughout that phase both for modes outside and also inside the Hubble radius, i.e $q/a>H$.  Since $\Phi$ is constant everywhere, no particle production occurs during reheating.
At a technical level this follows from the fact that if $\Phi$ is constant, integrating over $\tau$ the Bogoliubov coefficients yields a term proportional to $\delta(k+\omega)$.
As a consequence, modes that re-enter the Hubble radius during reheating will produce particles at the time $\tau_R$ corresponding to the reheating temperature. 
As it turns out this boosts the abundance because the particles produced are not diluted by entropy production. 

In formulae, the transfer function is constant to its inflationary value with $k_{\rm matter}=-3/5$, until reheating and then it is the solution of
\begin{equation}
T''(q,\tau)+\frac 4 {\tau} T'(q,\tau) + \frac{q^2}3 T(q,\tau)=0\,,\quad T(q,\tau_R)=-\frac{2}{3} \,,\quad T'(q,\tau_R)=0
\end{equation}
This matching injects a discontinuity in the transfer function at $\tau_R$, where it jumps from $-3/5$ to $-2/3$. This can be included in a smooth way, but it does not play a big role for our discussion, therefore for the sake of simplicity we stick to this approach.
The full transfer function in the approximation of finite reheating phase and sudden transition from matter to radiation is then
{\small
\begin{equation}
\begin{split}
T(q,\tau)&=-\frac{2 \left(q \left(9 \tau_R+\tau  \left(\tau_R^2 q^2-9\right)\right) \cos \left(\frac{q (\tau -\tau_R)}{\sqrt{3}}\right)+\sqrt{3} \left(\tau_R q^2 (3 \tau -\tau_R)+9\right) \sin \left(\frac{q (\tau -\tau_R)}{\sqrt{3}}\right)\right)}{3 \tau ^3 q^3}\theta(\tau-\tau_R)\\
&-\frac 3 5\theta(\tau_R-\tau)
\end{split}
\ee}

With this transfer function we can compute $A^\zeta_{\rm reh}$ in (\ref{eq:Azeta}) that determines the abundance.  

Importantly, the result depends very weakly from the reheating time $\tau_R$. This can be understood because particle production is simply delayed 
at reheating where potential start to evolve as during instantaneous reheating. Given that particle production depends on the non-adiabaticity, such a result might be expected. 
In Fig. \ref{fig:transfer} right we show the $A^\zeta$ during radiation and reheating as function of $\tau_R$. 
In the latter case the result follows the oscillations of the transfer function with a value that is slightly smaller than the result in radiation. 
In Fig. \ref{fig:abdiff} we show the differential abundance of single particles and of pairs for Weyl fermions. For the primordial power spectrum we have considered a monochromatic case where $\Delta_\zeta(q)\propto \delta(q-q_*)$.
Note that the first mildly depends on the spin while the latter differs only for the normalization proportional to the central charge. 

\begin{figure}
\begin{center}
\includegraphics[width=0.45\textwidth]{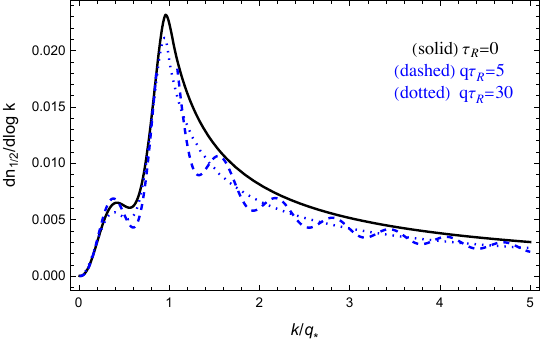}\quad \quad
\includegraphics[width=0.45\textwidth]{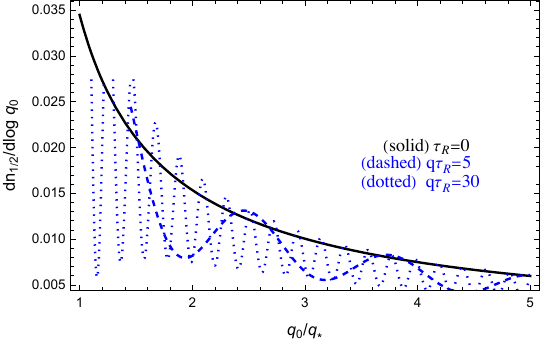} \\
\caption{\label{fig:abdiff} \textit{Left: Differential abundance of fermions produced by scalar inhomogeneities for momenta that re-enter the horizon during radiation (black) or reheating (blue dashed $q \tau_R=5$,
blue dotted $q \tau_R=30$).
Right: Differential abundance of particle pairs produced as function of total energy. 
The scalar power spectrum is taken to be proportional to $\delta(q-q_{\rm *})$.
Arbitrary units on the y axis.}}
\end{center}

\end{figure}

\paragraph{Application: DM from scalar perturbations.}~\\
What are the implications of this mechanism? An obvious application is DM production. Assuming DM is massless at the time of production, we have that the numerical abundance of DM is fully determined by the power spectrum of primordial scalar fluctuations, up to an overall factor that slightly depends on cosmology. See \cite{Garani:2024isu} for more details.

If the primordial spectrum $\Delta_\zeta(q)$ is not sharply peaked, then the integral is dominated by the momenta that exited the horizon at the end of
inflation, $q_{\rm max} \approx 10^{-7}$ eV. Instead, if the power spectrum is sharply peaked, $q_{\rm max}$ should be replaced with $q_{\rm peak}$.
An approximate formula is given by
\begin{equation}
na^3 \approx 10^{-4}\, \times c_J \Delta(q_{\rm peak}) q_{\rm peak}^3\,.
\end{equation}
Assuming that the particles produced have mass $M$ the abundance today  is given by,
\begin{equation}
\Omega_{\rm DM}\equiv  \frac {Mn a_0^3}{3 M_{\rm Pl}^2 H_0^2} \approx \frac {c_J\,\Delta(q_{\rm peak})}{10^{-2}}  \left(\frac{q_{\rm peak}}{10^{-7}\, {\rm eV}}\right)^3\frac {M}{10^7\,{\rm GeV}}\,.
\end{equation}
If the sector is interacting, for example it is made of a dark gauge theory that undergoes a confinement phase transition,
the abundance will be modified by order one factors that depends on the details of the dynamics. 
We leave a phenomenological analysis to future works.

\subsection{Tensor perturbations}

During inflation gravity waves are also produced through the same mechanism that gives rise to scalar perturbations.
Indeed the massless graviton satisfies the same equations of a minimally coupled scalar field. 
Contrary to scalar perturbations however tensor perturbations $\Delta_T$ are expected to be small. Bounds on $B$ modes in the CMB 
imply $H_I< 10^{14}$ GeV for the modes that exit the horizon during the first e-foldings of cosmological inflation. Since $H_I$ must be
decreasing in ordinary inflationary scenarios it follows that $\Delta_h < 10^{-10}$.

Despite the smallness of the power spectrum it is however interesting to compute the abundance produced because as this shows a new
phenomenon compared to scalar perturbations. For gravity waves again the matching is done on super-horizon scale to the tensor fluctuations of the metric.
Even in this case we can write
\be
\Delta_h(q,q_0)= \Delta_T (q)\times |\mathcal I_h (q,q_0)|^2,\quad\quad \text{with}\quad \quad\mathcal{I}_h(q\,,q_0)\equiv  \int_{-\infty}^\infty d\tau \,e^{-iq_0\tau}T_h(q,\tau)\,,
\ee
where $T_h(q,\tau)$ is the transfer function for gravity waves. In our notation, the transfer function satisfies
\be
T_h''+ 2 \frac{a'}{a}T'+ q^2 T_h=0,\quad \quad T_h(q,0)=1\,,\quad T'_h(q,0)=0\,.
\ee
At leading order the transfer function just depends on the scale factor $a(\tau)$ and not on the dynamics of the fluctuations of other species, as in general happens for the scalar fluctuations.

\bigskip

Let us discuss the case of instantaneous reheating. Here $\tau=0$ is a transition between inflation and radiation domination. In the radiation phase the transfer function reads
\be
T_h(q,\tau)=\theta(\tau) \frac{\sin (q \tau)}{q\tau}\,\quad \text{in radiation}\,.
\ee
If we are willing to compute the Fourier transform of this function, one gets
\begin{equation}
\mathcal{I}_h(q,q_0)\equiv  \int_0^\infty d\tau e^{-i q_0 \tau}\theta(\tau) \frac{\sin (q \tau)}{q\tau} =-\frac{i}{q_0} \tanh ^{-1}\left(\frac{q}{q_0}\right)= - \frac i {q_0} +\dots 
\end{equation}
Since the gravitational kernels grow as $k^2$ (instead of $q^4/k^2$ for scalar perturbations, cfr eq.s~\eqref{eq:tensor-kernels}) one naively finds that the abundance is cubically divergent. This conclusion is however an artefact of assuming a discontinuous gravity wave signal $\propto \theta(\tau)$ that leads to unphysical infinite production of particles.
In reality the transfer function is non zero during inflation. Since the modes are constant outside the horizon,
\begin{equation}
T_h(q\,,\tau) = \theta(-\tau)\quad\text{in inflation}\,,
\end{equation}
so that the full Fourier transform is given by
\begin{equation}
 \mathcal{I}_h(q,q_0)= \frac{i}{q_0}-\frac{i \tanh ^{-1}\left(q/q_0\right)}{q}=-\frac{i q^2}{3 q_0^3} +\dots
\end{equation}
In this way the result is actually finite.

This result is actually obvious from the fact that the leading term is proportional to the Weyl tensor of the background. 
Any smooth background that goes sufficiently fast to zero at infinity produces a finite result. It is however interesting that the abundance
depends on the UV assumptions. 

The gravitational induced particle production from inflation thus take the form
\begin{equation}
na^3 =c_J \frac  {A^{\rm GW}}{4\pi^2}  \int d (\log q) q^3 \Delta_T (q)\,,\quad\quad  A^{\rm GW}=   \frac 1 {160}\int_1^{\infty} dx  |\mathcal{I}_h(1,x)|^2\,.
\end{equation}
For modes that re-enter the horizon during radiation domination $A^{\rm GW}_{\rm rad}=0.0005$.

\section{Conclusions and outlook}
\label{sec:conclusions}

In this work we computed in full generality the production of conformally coupled theories from metric inhomogeneities.
This applies in particular to massless fermions, gauge fields and conformally coupled scalars that can be produced from scalar perturbations
and gravity waves. The formalism also captures interacting CFTs whose production is determined by the central charge of the theory. 

\medskip

Our formalism provides a unified framework for many results previously derived in the literature. 
The inclusive production can be determined most simply using the 1PI effective action in the presence of the background,
similarly to Schwinger pair production. This derivation only requires  the knowledge of the 2-point function of the energy momentum 
tensor that is  fixed for conformally coupled theories up to the central charge. The final result is proportional to
the square of the Weyl tensor in momentum space. More exclusive information on the distribution of particles can be determined computing Bogoliubov coefficients
that require the knowledge of the lagrangian. We presented explicit formulas for conformally coupled scalars, fermions and gauge fields. 
Interestingly the Bogoliubov coefficients can be cast as on-shell helicity amplitudes in the background of metric inhomogeneities. We believe that this deserves further study in the future, especially with the aim of applying the same techniques to massive theories.

\medskip

We  envision several phenomenological applications. 

One direction is to quantify the abundance of DM produced from scalar or tensor perturbations in different dark sectors. Whenever inhomogeneities  are active, particles are created even when they are massless. For example, in \cite{Garani:2024isu} we explored the case of scalar perturbations produced during inflation finding that the cosmological abundance of DM can be generated if the power spectrum is enhanced towards the end of inflation, leading to a strong connection between the origin of DM and the physics at the end of inflation. Since our formalism applies in general to CFTs, one can consider the production of interacting dark sectors, which can host interesting DM candidates and cosmological signals. Among them, we are planning to quantify the impact of our mechanism on the production of dark gauge theories (theories of DM glueballs) or even interacting CFTs (possibly with their holographic dual). 

Phase transitions are also a natural arena where we expect our results to be relevant. 
Large effects for metric perturbations are in particular expected  in cosmological first order phase transition and second order if associated to the formation of topological defects.

Finally, beside cosmological backgrounds our results allow to study the production of particles from localized sources. 
We are planning to study such effects in astrophysical environment and their interplay with Hawking radiation.

\subsubsection*{Acknowledgements}
{\small We would like to thank Marco Peloso for discussions on Bogoliubov transformations.
The work of AT and MR is supported by the European Union – Next Generation EU and by the
Italian Ministry of University and Research (MUR) via the PRIN 2022 project n. 20228WHTYC.}

\appendix

\section{More on Bogoliubov transformations}\label{app:bogo}

In this appendix we provide technical details of our computations.

\subsection{Particle number and energy}
To determine the particles produced the relevant quantity is the number operator per mode, 
written in terms of the "out" basis, $\bar a_k^\dag \bar a_k$, as this turns out important to compute observables, such as number of particles produced and the energy associated to them.
From eq. (\ref{eq:bogorotation}) the particle number in the mode $\vec k$ can be expressed in terms of the initial basis as
\begin{eqnarray}
\lin \bar a^\dag_{ \vec k} \bar a_{ \vec k} \rin & =& \int \frac{d^3q}{(2\pi)^3}\int \frac{d^3q'}{(2\pi)^3}\lin  (\beta_{\vec k, -\vec \omega} a_{\vec \omega}+ \alpha^*_{\vec k, -\vec \omega}a^\dag_{ -\vec \omega})(\alpha_{\vec k, \vec \omega'} a_{\vec \omega'}+ \beta^*_{\vec k, \vec \omega'}a^\dag_{ -\vec \omega'}) \rin\nonumber\\
&=& \int \frac{d^3q}{(2\pi)^3}\int \frac{d^3q'}{(2\pi)^3}\, \beta_{\vec k, -\vec \omega} \beta^*_{\vec k, \vec \omega'} \lin a_{ \vec \omega} a^\dag_{ -\vec \omega'} \rin = \int \frac{d^3q}{(2\pi)^3}\, |\beta_{\vec k, \vec \omega}|^2
\end{eqnarray}
Integrating over $\vec k$ we obtain \eqref{eq:Ndiff0}. We can therefore compute the energy as
\begin{eqnarray}
\lin H^{\rm out} \rin &=& \int \frac{d^3k}{(2\pi)^3} k \lin \bar a^\dag_{ \vec k} \bar a_{ \vec k} \rin  + (2\pi)^3\delta(0)\int \frac{d^3k}{(2\pi)^3} \frac{k}{2} 
\end{eqnarray}
where the last term is the vacuum energy.

\subsection{Equivalence between Schwinger and Bogoliubov}\label{sec:equivalence}

In this appendix we sketch the proof of the equivalence between the computation using the vacuum persistence 
and particle production from Bogoliubov coefficients. We follow the discussion in \cite{Ford:2021syk} and in \cite{Parker:1969au}
generalizing it to inhomogeneous background.  As discussed in section \ref{sec:inout}, the vacuum to vacuum transition amplitude in terms of effective action is exactly given by, 
\begin{equation}
   \langle {\rm out} \rin  = e^{i\Gamma}~.
\end{equation}
It is possible to evaluate the same overlap $\langle {\rm out} \rin  $ using Bogoliubov coefficients. Inverting the relation in \eqref{eq:bogorotation} we can express the ``in'' operators in terms of the ``out'' ones. Working to leading order in perturbations,
\begin{equation}
    a_{\vec{k}} =\bar{a}_{ \vec{k}}-\int \frac {d^3 q}{(2\pi)^3} \left[\delta \alpha_{\vec{k},\vec{\omega}} \,\bar{a}_{ \vec{\omega}}+ \beta^\ast_{\vec{k}\vec{\omega}} \,\bar{a}^\dag_{-\vec{\omega}}\right]\,.
\end{equation}
The idea is to construct the state $\rin$ acting on $\rout$ with the operators ``out".
One can check that the state,
\begin{equation}
| v\rangle=\Pi_{\vec k}\left[1+ \frac {1} {(2\pi)^3 \delta^3(0)} \int \frac{d^3 q}{(2\pi)^3}\beta_{\vec k \vec \omega} \bar{a}^\dag_{\vec{k}} \bar{a}^\dag_{ -\vec{\omega}}\right] \rout
\end{equation}
is annihilated by $a_{\vec{k}}$ to first order in the perturbation and for each $\vec k$. As a consequence it must be proportional to the $\rin$.
We then find,
\begin{equation}
   \langle {\rm out} \rin  \approx \frac {   \langle {\rm out} | v\rangle }{\sqrt{\langle v| v\rangle}}\approx 1 -\frac 1 2\int \frac {d^3k}{(2\pi)^3}\int \frac{d^3q}{(2\pi)^3} |\beta_{\vec k \vec \omega} |^2
\end{equation}
Comparing with eq. (\ref{persistence}) we find,
\begin{equation}
2 \mathrm{Im}\Gamma= \int \frac{d^3k}{(2\pi)^3}\int \frac{d^3q}{(2\pi)^3} |\beta_{\vec k \vec \omega}|^2
\end{equation}
An extra factor 1/2 must be included for identical particles.
\section{Tensors and polarizations}
\label{app:tensors}

In this appendix we collect useful formulae for the tensors that appear in the paper. 
We consider perturbations of the FLRW metric eq.~\eqref{eq:metric}, $ds^2= a^2(\tau)[ \eta_{\mu\nu}+ h_{\mu\nu}(\tau\,, \vec x)] dx^\mu dx^\nu$.
We evaluate all relevant geometric quantities up to first order in metric perturbations below, with the co-moving Hubble parameter $\mathcal{H} = a^\prime/a$.

\paragraph{Riemann tensor:}~\\
The perturbed Riemann tensor reads
\begin{equation}
   R_{\mu \nu \rho \sigma}=   R^{(0)}_{\mu \nu \rho \sigma} + R_{\mu \nu \rho \sigma}^{(1)} ~.
\end{equation}

The background Riemann tensor is 
\begin{equation}
    \frac{1}{a^2}R^{(0)}_{\mu \nu \rho \sigma} =   \mathcal{H}^\prime \delta_{ij} \lp \delta_{\rho 0} \delta^j_\sigma -  \delta^j_\rho \delta_{\sigma 0} \rp\delta_{\mu0} \delta^i_\nu + \mathcal{H}^2 \delta_{ij} \delta_{kl} \lp \delta^k_\nu \delta^j_\rho -  \delta^j_\nu \delta^k_\rho   \rp \delta^i_\mu \delta^l_\sigma.
\end{equation}
The perturbed contribution reads

\begin{eqnarray}
    \frac{2}{a^2} ( R_{\mu \nu \rho \sigma}^{(1)}) &=& - 2 \mathcal{H}^\prime h_{ij} \lp \delta_{\rho0} \delta^j_\sigma -  \delta^j_\rho \delta_{\sigma0} \rp\delta_{\mu0} \delta^i_\nu  - \mathcal{H} \lp \delta_{i j} \partial_0 h_{00}  + \partial_0 h_{i j} - \partial_i h_{0 j} - \partial_j h_{0 i} \rp \delta_{\mu0} \delta^i_\nu  \delta_{\rho 0} \delta^j_\sigma \nonumber \\
    & & + \delta_{i j} \lp -2 \mathcal{H}^2  h_{0 k} + \mathcal{H}  \partial_k h_{0 0} - \partial_i \partial_k h_{0 j} + \partial_i \partial_j h_{0 k} + \partial_0 \partial_k h_{i j}  \rp \delta_{\mu0} \delta^i_\nu  \delta^j_\rho \delta^k_\sigma  \nonumber \\
    && + \delta_{ik} \lp 2\mathcal{H}^2 h_{j l} - \mathcal{H}  \lp -\partial_0 h_{j l} + \partial_l h_{0 j} +\partial_j h_{0 l} \rp  - \partial_j \partial_l h_{ik} - \partial_i \partial_k h_{jl} \rp\delta^i_\mu \delta^j_\nu  \delta^k_\rho \delta^l_\sigma \nonumber \\
    & & +   \partial_\nu \partial_\rho h_{\mu \sigma} - \partial_\nu \partial_\sigma h_{\mu \rho}  + \partial_\sigma \partial_\mu h_{\rho \nu} - \partial_\rho \partial_\mu h_{\sigma \nu}~.
\end{eqnarray}

\paragraph{Ricci tensor:}~\\ 
Contracting the Riemann tensor we get

\begin{equation}
   R_{\mu \nu }=   R^{(0)}_{\mu \nu } +  R_{\mu \nu}^{(1)} ~,
\end{equation}
with $h = g^{\mu \nu} h_{\mu\nu}$. Explicitly we have the following
\begin{equation}
  R^{(0)}_{\mu \nu } = - 3 \mathcal{H}^\prime \delta_{\mu 0} \delta_{\nu 0} + \delta_{ij}\lp \mathcal{H}^\prime + 2 \mathcal{H}^2\rp \delta^i_\mu\delta^j_\nu~,
\end{equation}

and

\begin{eqnarray}
 2  R_{\mu \nu }^{(1)} &=&  \left( \mathcal{H} \left( 3 \partial_0 h_{00} + \partial_0 h_{ii} - 2 \partial_i h_{0i} \right) +   \nabla^2 h_{00} + \partial_0^2 h_{ii}  - 2 \partial_0 \partial_i h_{0 i} \right) \delta_{\mu 0} \delta_{\nu 0}  \nonumber \\
 && -  \left( 2 h_{0i} \left( 2 \mathcal{H}^2 + \mathcal{H}^\prime \right) - 2 \mathcal{H} \,\partial_i h_{00}    - \nabla^2 h_{0i} + \partial_0 \partial_j h_{i j}-  \partial_0 \partial_i h_{j j}  + \partial_i \partial_j h_{0 j} \right) \delta_{\mu 0} \delta^i_{\nu} \nonumber \\
 && - \left( 4 \mathcal{H}^2 + 2  \mathcal{H}^\prime \right) \left( \delta_{ij} h_{00}+  h_{i j} \right) \delta^i_\mu \delta^j_\nu \nonumber \\
 && + 2 \mathcal{H} \left( \delta_{i j} \frac{1}{2} \partial_0 h_{00}  - \delta_{ij} \frac{1}{2} \partial_0 h_{kk} + \delta_{ij} \partial_k h_{0 k}  - \partial_0 h_{i j} + \partial_j h_{0i} + \partial_i h_{0j} \right) \delta^i_\mu \delta^j_\nu  \nonumber \\
 && + \partial_\lambda \partial_\nu h^\lambda_\mu + \partial^\lambda \partial_\mu h_{\nu \lambda} - \Box h_{\mu \nu} - \partial_\nu \partial_\mu h~.
\end{eqnarray}

\paragraph{Ricci scalar:} The perturbed Ricci scalar has the form

\begin{equation}
   R=   R^{(0)} +  R^{(1)} ~.
\end{equation}

The background Ricci tensor is 
\begin{equation}
  a^2 R^{(0)}=  - 6 \lp \mathcal{H}^\prime  + \mathcal{H}^2  \rp~,
\end{equation}
\begin{eqnarray}
a^2   R^{(1)} &=& 6 h_{00} \lp \mathcal{H}^2 + \mathcal{H}^\prime\rp  + 3 \mathcal{H} \lp \partial_0 h_{00} + \partial_0 h_{ii} - 2 \partial_i h_{0i}\rp  +  \partial^\mu \partial^\nu h_{\mu \nu} -  \Box h~.
\end{eqnarray}


\paragraph{Weyl tensor:}~\\ 
The trace free part of the Riemann tensor is the Weyl tensor. In d ($\geq 3$) dimensions the following expression holds, 
\begin{equation}
    W_{\mu \nu \rho \sigma} = R_{\mu \nu \rho \sigma} - \frac{2}{d-2} \lp g_{\mu[\rho} R_{\sigma] \nu} -g_{\nu[\rho} R_{\sigma] \mu} \rp + \frac{2}{(d-1)(d-2)} R g_{\mu[\rho}g_{\sigma] \nu}~.
\end{equation}
The invariant is independent of background expansion and only depends on the geometry through the derivatives of perturbations. We can write
\begin{equation}
   W_{\mu \nu \rho \sigma}=   W_{\mu \nu \rho \sigma}^{(0)} +  W_{\mu \nu \rho \sigma}^{(1)} ~.
\end{equation}
The unperturbed part is

\begin{equation}
   W^{(0)}_{\mu \nu \rho \sigma}=   R^{(0)}_{\mu \nu \rho \sigma} - \lp \bar{\eta}_{\mu[\rho]} R^{(0)}_{\sigma]\nu} - \bar{\eta}_{\nu[\rho} R^{(0)}_{\sigma]\mu}  \rp  + \frac{1}{3} R^{(0)} \bar{\eta}_{\mu[\rho}\bar{\eta}_{\sigma]\nu}  = 0~,
\end{equation}
with $\bar{h}_{\mu \nu} = a^2 h_{\mu \nu} $ and $\bar{\eta}_{\mu \nu} = a^2 \eta_{\mu \nu} $. 
The inhomogeneous part is given by

\begin{eqnarray}
 W_{\mu \nu \rho \sigma}^{(1)} &=&    R_{\mu \nu \rho \sigma}^{(1)} - \lp \bar{h}_{\mu[\rho]} R^{(0)}_{\sigma]\nu} - \bar{h}_{\nu[\rho} R^{(0)}_{\sigma]\mu}  + 
 \bar{\eta}_{\mu[\rho]}  R_{\sigma]\nu}^{(1)} - \bar{\eta}_{\nu[\rho}  R_{\sigma]\mu}^{(1)}  \rp  \nonumber \\
 && + \frac{1}{3} \lp  R^{(1)} \bar{\eta}_{\mu[\rho}\bar{\eta}_{\sigma]\nu} + R^{(0)} \bar{h}_{\mu[\rho}\bar{\eta}_{\sigma]\nu} + R^{(0)} \bar{\eta}_{\mu[\rho}\bar{h}_{\sigma]\nu} \rp~.
\end{eqnarray}
The Weyl tensor squared is
\begin{eqnarray}
a^4 W^2 &=&
\frac{1}{4} (\partial_\mu \partial_\sigma h_{\rho\nu}) (\partial^\mu \partial^\sigma h^{\rho\nu}) -  \frac{1}{3} (\partial^\mu \partial^\nu h_{\mu\nu})^2 - \frac{2}{3} (\partial^\mu \partial^\nu h_{\mu\nu}) (\Box h) + \frac{1}{3} (\Box h)^2~.
\end{eqnarray}

\paragraph{Spin connection:}~\\ 
Dealing with fermions requires the introduction of vierbeins, defined as 
\begin{equation}
    g_{\mu \nu} = \textbf{e}^a_\mu \textbf{e}^b_\nu \eta_{a b} ~.
\end{equation}
In the presence of perturbations $g_{\mu \nu} = a^2\lp\eta_{\mu \nu} + h_{\mu\nu}\rp$, up to linear order we have explicitly the following
\begin{eqnarray}
   \textbf{e}^a_\mu &=& a(\tau) (\delta^a_\mu + \frac{1}{2} \delta^{a \nu} h_{\mu\nu}) \quad \equiv \quad  \textbf{e}^a_\mu + \delta \textbf{e}^a_\mu~.
\end{eqnarray}
For fermions the covariant derivative is given by
\begin{equation}
  D_\mu = \partial_\mu + \frac{1}{4}\omega^{a b}_{\mu} \gamma_{[a} \gamma_{b]}~.
  \end{equation}
The spin connection defined in terms of the above tetrads reads
\begin{equation}
       \omega^{ab}_\mu = 2 \te^{\nu[a} \partial_{[ \mu}   \te^{b]}_{\nu]} - \te^{\nu[a} \te^{b]\sigma} \te_{\mu c} \partial_{\nu} \te^c_\sigma~.
\end{equation}
For the line element used in this work (FRW, eq.~\eqref{eq:metric}) this is given by 
\begin{equation}
    \omega_\mu^{ab} = \mathcal{H} \Bigg(\lp\delta^{b0} \delta^a_\mu - \delta^{a0} \delta^b_\mu \rp + \frac{1}{2} \delta_{\alpha 0} \lp h^{\alpha a} \delta^b_\mu - h^{\alpha b} \delta^a_\mu \rp + \frac{1}{2} \lp\delta^{b0} h^a_\mu -  \delta^{a0} h^b_\mu \rp\Bigg) - \frac{1}{2} \lp \partial^a h^b_\mu - \partial^b h^a_\mu \rp~.
\end{equation}

\subsection{Energy momentum tensor}
In the paper we derived the equations of motion to first order in perturbations expanding the equations of motions in curved space.
An alternative way is to expand the action to first order in perturbations,
\begin{equation}
S= S_0 +\frac 1 2 \int d^4x h_{\mu\nu}(x)T^{\mu\nu}(x)
\end{equation}
and derive the equations of motion. 

Around flat space we have,
\begin{equation}
\begin{split}
T^{\mu\nu}_{\rm scalar} &= \partial^\mu \varphi \partial^\nu \varphi - \frac{1}{2}\eta^{\mu\nu} \partial^\lambda \varphi \partial_\lambda \varphi + \xi \lp \eta^{\mu \nu} \Box -\partial^\mu \partial^\nu \rp \varphi^2 \nonumber \\
 T^{\mu\nu}_{\rm Weyl} &= \frac{1}{4} i \lp \bar{\psi} \bar{\sigma}^{(\mu} \partial^{\nu)} \psi -  \partial^{(\mu} \bar{\psi} \bar{\sigma}^{\nu)}\psi \rp~- i  \eta^{\mu\nu}\bar{\psi}\bar{\sigma}^\mu \partial_\mu \psi,  \nonumber \\
T^{\mu\nu}_{\rm gauge}& = F^{\mu \rho}F_\rho^{\,\,\nu} + \frac14 \eta^{\mu\nu} F^2~.
\end{split}
\end{equation}
For conformally coupled scalars considered in the paper $\xi=1/6$.

\subsection{Polarizations}\label{appC}

Our results in general depend on the polarization of particles and also of the helicity structure of the gravitational background.

For the particles we work in the helicity basis. Here, using as reference a four-vector $p^\mu=(|\vec p|, \vec p)$, the polarization vector and/or spinors are eigenstates of the operator $\vec p \cdot \vec J$ with eigenvalues $\pm |\vec p|$.

\begin{itemize}
\item For fermions, the two polarizations spinors $\xi_{\vec p}^\pm$ are solutions of the Dirac equation with positive/negative frequencies
\be
\bar\sigma^\mu p_\mu \xi_{\vec p}^-=0\,,\quad \sigma^\mu p_\mu \xi_{\vec p}^+=0\,.
\ee
\item For vectors, in Coloumb gauge, the two polarization vectors $\varepsilon_\mu^\pm(p)=(0,\vec\varepsilon^\pm)$ are solutions of the equations
\be
\vec p\cdot \vec\varepsilon^\pm =0\,\quad (\vec p \cdot \vec J_p)  \varepsilon^{\pm}=\pm |p|  \varepsilon^{\pm}
\ee
where $\vec J_p$ is the generator of rotation around $\vec p$.
\end{itemize}

For gravitational waves, that in our context are always labelled by vector $\vec q$, we use the helicity basis. 
For $\vec q$ aligned along the $\hat z$ axis (see \cite{Weinberg:2008zzc}) the polarization tensors can be chosen,
\be
\small \epsilon_{3}^\pm = \frac{1}{\sqrt{2}}\begin{pmatrix} 
1 & \pm i & 0 \\
\pm i & -1 & 0 \\
0 & 0 & 0
\end{pmatrix}
\ee
In a frame where $\vec q= q(\sin \theta \cos \phi\,,\sin \theta \sin \phi\,, \cos \theta)$, the polarization are then obtained  performing the  rotation  $\vec q = q R(\theta,\phi) \hat z $.
Explicitly
\be
\epsilon_{ij}^\pm(\vec q)  = R_{ik}(\theta,\phi)R_{jl}(\theta,\phi) (\epsilon_{3}^\pm)_{kl} = [ R \cdot \epsilon_{3}^\pm\cdot R^T ]_{ij}\,\,,
\ee
where
\begin{equation}
R(\theta,\phi)=\left(
\begin{array}{ccc}
 \cos (\theta ) \cos (\phi ) & -\sin (\phi ) & \sin (\theta ) \cos (\phi ) \\
 \cos (\theta ) \sin (\phi ) & \cos (\phi ) & \sin (\theta ) \sin (\phi ) \\
 -\sin (\theta ) & 0 & \cos (\theta ) \\
\end{array}
\right)\,.
\end{equation}

\pagestyle{plain}
\bibliographystyle{jhep}
\small
\bibliography{biblio_stochastic_long}

\providecommand{\href}[2]{#2}\begingroup\raggedright\begin{thebibliography}{10}

\bibitem{Ford:1986sy}
L.~H. Ford, {\it {Gravitational Particle Creation and Inflation}},  {\em Phys.
  Rev. D} {\bf 35} (1987) 2955.

\bibitem{Kolb:2023ydq}
E.~W. Kolb and A.~J. Long, {\it {Cosmological gravitational particle production
  and its implications for cosmological relics}},  {\em Rev. Mod. Phys.} {\bf
  96} (2024), no.~4 045005, [\href{http://arxiv.org/abs/2312.09042}{{\tt
  arXiv:2312.09042}}].

\bibitem{kopp}
A.~Maleknejad and J.~Kopp, {\it {Gravitational Wave-Induced Freeze-In of
  Fermionic Dark Matter}},  \href{http://arxiv.org/abs/2405.09723}{{\tt
  arXiv:2405.09723}}.

\bibitem{kopp2}
A.~Maleknejad and J.~Kopp, {\it {Weyl fermion creation by cosmological
  gravitational wave background at 1-loop}},  {\em JHEP} {\bf 01} (2025) 023,
  [\href{http://arxiv.org/abs/2406.01534}{{\tt arXiv:2406.01534}}].

\bibitem{Garani:2024isu}
R.~Garani, M.~Redi, and A.~Tesi, {\it {Stochastic Dark Matter from Curvature
  Perturbations}},  \href{http://arxiv.org/abs/2408.15987}{{\tt
  arXiv:2408.15987}}.

\bibitem{Redi:2020ffc}
M.~Redi, A.~Tesi, and H.~Tillim, {\it {Gravitational Production of a Conformal
  Dark Sector}},  {\em JHEP} {\bf 05} (2021) 010,
  [\href{http://arxiv.org/abs/2011.10565}{{\tt arXiv:2011.10565}}].

\bibitem{Zeldovich:1977vgo}
Y.~B. Zel'dovich and A.~A. Starobinsky, {\it {Rate of particle production in
  gravitational fields}},  {\em JETP Lett.} {\bf 26} (1977), no.~5 252.

\bibitem{Calzetta:1986ey}
E.~Calzetta and B.~L. Hu, {\it {Closed Time Path Functional Formalism in Curved
  Space-Time: Application to Cosmological Back Reaction Problems}},  {\em Phys.
  Rev. D} {\bf 35} (1987) 495.

\bibitem{Campos:1991ff}
A.~Campos and E.~Verdaguer, {\it {Production of spin 1/2 particles in
  inhomogeneous cosmologies}},  {\em Phys. Rev. D} {\bf 45} (1992) 4428--4438.

\bibitem{Cespedes:1989kh}
J.~Cespedes and E.~Verdaguer, {\it {Particle Production in Inhomogeneous
  Cosmologies}},  {\em Phys. Rev. D} {\bf 41} (1990) 1022.

\bibitem{Maroto1}
A.~L. Maroto, {\it {Primordial magnetic fields from metric perturbations}},
  {\em Phys. Rev. D} {\bf 64} (2001) 083006,
  [\href{http://arxiv.org/abs/hep-ph/0008288}{{\tt hep-ph/0008288}}].

\bibitem{Maroto2}
A.~L. Maroto, {\it {Constraining the primordial spectrum of metric
  perturbations from gravitino and moduli production}},  {\em Phys. Rev. D}
  {\bf 65} (2002) 083508, [\href{http://arxiv.org/abs/hep-ph/0111126}{{\tt
  hep-ph/0111126}}].

\bibitem{Hu:2020luk}
B.-L.~B. Hu and E.~Verdaguer, {\em {Semiclassical and Stochastic Gravity}:
  {Quantum Field Effects on Curved Spacetime}}.
\newblock Cambridge Monographs on Mathematical Physics. Cambridge University
  Press, Cambridge, 1, 2020.

\bibitem{Bassett:2001jg}
B.~A. Bassett, M.~Peloso, L.~Sorbo, and S.~Tsujikawa, {\it {Fermion production
  from preheating amplified metric perturbations}},  {\em Nucl. Phys. B} {\bf
  622} (2002) 393--415, [\href{http://arxiv.org/abs/hep-ph/0109176}{{\tt
  hep-ph/0109176}}].

\bibitem{Schwinger:1951nm}
J.~S. Schwinger, {\it {On gauge invariance and vacuum polarization}},  {\em
  Phys. Rev.} {\bf 82} (1951) 664--679.

\bibitem{Watkins:1991zt}
R.~Watkins and L.~M. Widrow, {\it {Aspects of reheating in first order
  inflation}},  {\em Nucl. Phys. B} {\bf 374} (1992) 446--468.

\bibitem{Gubser:1997se}
S.~S. Gubser and I.~R. Klebanov, {\it {Absorption by branes and Schwinger terms
  in the world volume theory}},  {\em Phys. Lett. B} {\bf 413} (1997) 41--48,
  [\href{http://arxiv.org/abs/hep-th/9708005}{{\tt hep-th/9708005}}].

\bibitem{Birrell:1982ix}
N.~D. Birrell and P.~C.~W. Davies, {\em {Quantum Fields in Curved Space}}.
\newblock Cambridge Monographs on Mathematical Physics. Cambridge University
  Press, Cambridge, UK, 1982.

\bibitem{Luty:2012ww}
M.~A. Luty, J.~Polchinski, and R.~Rattazzi, {\it {The $a$-theorem and the
  Asymptotics of 4D Quantum Field Theory}},  {\em JHEP} {\bf 01} (2013) 152,
  [\href{http://arxiv.org/abs/1204.5221}{{\tt arXiv:1204.5221}}].

\bibitem{Ford:2021syk}
L.~H. Ford, {\it {Cosmological particle production: a review}},  {\em Rept.
  Prog. Phys.} {\bf 84} (2021), no.~11
  [\href{http://arxiv.org/abs/2112.02444}{{\tt arXiv:2112.02444}}].

\bibitem{Osborn:1993cr}
H.~Osborn and A.~C. Petkou, {\it {Implications of conformal invariance in field
  theories for general dimensions}},  {\em Annals Phys.} {\bf 231} (1994)
  311--362, [\href{http://arxiv.org/abs/hep-th/9307010}{{\tt hep-th/9307010}}].

\bibitem{Weinberg:2008zzc}
S.~Weinberg, {\em {Cosmology}}.
\newblock 2008.

\bibitem{Parker_Toms_2009}
L.~Parker and D.~Toms, {\em Quantum Field Theory in Curved Spacetime: Quantized
  Fields and Gravity}.
\newblock Cambridge Monographs on Mathematical Physics. Cambridge University
  Press, 2009.

\bibitem{Parker:1969au}
L.~Parker, {\it {Quantized fields and particle creation in expanding universes.
  1.}},  {\em Phys. Rev.} {\bf 183} (1969) 1057--1068.

\end{thebibliography}\endgroup

\end{document}